%% file: MVASLAMbounds.tex
\documentclass[conference]{IEEEtran}
\IEEEoverridecommandlockouts
\usepackage{ifthen}
\newcommand{\exportFigures}{false} 

\input{./definitions.tex}			

\acrodef{surface}[SFV]{surface feature vector}	
\newcommand{\surface}[0]{sfv}

\acrodef{pcrlb}[PCRLB]{posterior \ac{crlb}} 
\acrodef{dmimo}[D-MIMO]{distributed multiple-input multiple-output} 
\acrodef{mpslam}[MP-SLAM]{multipath-based simultaneous localization and mapping} 
\acrodef{mva}[MVA]{master virtual anchor} 

\acrodef{dl}[DL]{downlink}
\acrodef{peb}[PEB]{position error bound}
\acrodef{veb}[VEB]{velocity error bound}
\acrodef{oeb}[OEB]{orientation error bound}
\acrodef{meb}[MEB]{mapping error bound}
\acrodef{pm}[PM]{physical mobile}
\acrodef{vm}[VM]{virtual mobile}
\acrodef{mc}[MC]{Monte Carlo}

\acrodef{wpt}[WPT]{wireless power transfer}
\acrodef{nlos}[NLoS]{non-LoS}
\acrodef{olos}[OLoS]{obstructed LoS}

\usepackage{cite}
\usepackage{amsmath,amssymb,amsfonts}
\usepackage{algorithmic}
\usepackage{graphicx}
\usepackage{textcomp}
\usepackage{xcolor}
\usepackage{orcidlink}

\usepackage{array, booktabs, xltabular}
\usepackage{colortbl}       
\usepackage{multirow}	
\usepackage{longtable}	
\usepackage{nicefrac}	
\usepackage{adjustbox}          
\usepackage{upgreek}          
\usepackage{siunitx}          
\usepackage{balance}          

\usepackage[T1]{fontenc}

\hypersetup{
    colorlinks=false,
    linkcolor=black,
    pdfborder={0 0 0}
    }
\input{./InputFiles/math-notation}
\DeclareMathOperator{\arctantwo}{arctan2}
\DeclareMathOperator{\tr}{tr}
\newcommand{\trp}{\mathrm{T}}					

\ifthenelse{\equal{\exportFigures}{false}}
{
   \newcommand{\tikzexternaldisable}{}				
   \newcommand{\tikzexternalenable}{}				
  }
{
  \usepgfplotslibrary{external}
  \usepackage{shellesc}
}

\newcommand{\lineref}[1]{(\tikzexternaldisable\ref{#1}\tikzexternalenable)}					

\newcolumntype{C}{@{\hskip 0.075cm}c@{\hskip 0.075cm}}
\newenvironment{bmatrixs}
  {\left[\begin{array}{*{20}{C}}}
  {\end{array}\right]}

\DeclareSymbolFont{sfletters}{OML}{cmbrm}{m}{it}

\DeclareMathSymbol{\salpha}{\mathord}{sfletters}{"0B}
\DeclareMathSymbol{\sbeta}{\mathord}{sfletters}{"0C}
\DeclareMathSymbol{\sgamma}{\mathord}{sfletters}{"0D}
\DeclareMathSymbol{\sdelta}{\mathord}{sfletters}{"0E}
\DeclareMathSymbol{\sepsilon}{\mathord}{sfletters}{"0F}
\DeclareMathSymbol{\szeta}{\mathord}{sfletters}{"10}
\DeclareMathSymbol{\seta}{\mathord}{sfletters}{"11}
\DeclareMathSymbol{\stheta}{\mathord}{sfletters}{"12}
\DeclareMathSymbol{\siota}{\mathord}{sfletters}{"13}
\DeclareMathSymbol{\skappa}{\mathord}{sfletters}{"14}
\DeclareMathSymbol{\slambda}{\mathord}{sfletters}{"15}
\DeclareMathSymbol{\smu}{\mathord}{sfletters}{"16}
\DeclareMathSymbol{\snu}{\mathord}{sfletters}{"17}
\DeclareMathSymbol{\sxi}{\mathord}{sfletters}{"18}
\DeclareMathSymbol{\spi}{\mathord}{sfletters}{"19}
\DeclareMathSymbol{\srho}{\mathord}{sfletters}{"1A}
\DeclareMathSymbol{\ssigma}{\mathord}{sfletters}{"1B}
\DeclareMathSymbol{\stau}{\mathord}{sfletters}{"1C}
\DeclareMathSymbol{\supsilon}{\mathord}{sfletters}{"1D}
\DeclareMathSymbol{\sphi}{\mathord}{sfletters}{"1E}
\DeclareMathSymbol{\schi}{\mathord}{sfletters}{"1F}
\DeclareMathSymbol{\spsi}{\mathord}{sfletters}{"20}
\DeclareMathSymbol{\somega}{\mathord}{sfletters}{"21}
\DeclareMathSymbol{\svarepsilon}{\mathord}{sfletters}{"22}
\DeclareMathSymbol{\svartheta}{\mathord}{sfletters}{"23}
\DeclareMathSymbol{\svarpi}{\mathord}{sfletters}{"24}
\DeclareMathSymbol{\svarrho}{\mathord}{sfletters}{"25}
\DeclareMathSymbol{\svarsigma}{\mathord}{sfletters}{"26}
\DeclareMathSymbol{\svarphi}{\mathord}{sfletters}{"27}

\renewcommand{\posAnchor}[1]{ \V{p}_{\mathrm{pa}}^{(#1)}  }     
\newcommand{\posVA}[3]{ \V{p}_{#2#3,\mathrm{va}}^{(#1)} }     
\newcommand{\realset}[2]{\mathbb{R}^{#1\rmv\times\rmv#2}}

\newcommand{\rotMn}[0]{\M{M}_{\scriptscriptstyle n}}                                       
\newcommand{\rangepRX}[4]{\grave{\bm{r}}_{\scriptscriptstyle#1,#2}^{\scriptscriptstyle\MVApair{#3}{#4}}} 
\renewcommand{\pmva}[1]{\V{p}_{#1,\mathrm{\surface}} }               
\newcommand{\pmvaHat}[1]{\hat{\V{p}}_{#1,\mathrm{\surface}} }               
\newcommand{\RVpmva}[1]{\RV{p}_{#1,\mathrm{\surface}} }        

\renewcommand{\house}[1]{ \M{H}_{\scriptscriptstyle #1} }   	
\newcommand{\dist}[2]{d_{#1#2,n}^{(j)}}								
\newcommand{\aoa}[2]{\varphi_{#1#2,n}^{(j)}}						
\newcommand{\aod}[2]{\phi_{#1#2,n}^{(j)}}							
\renewcommand{\delayx}[1]{d_{ss',n}^{(#1)}}							
\renewcommand{\delay}[0]{d}												
\renewcommand{\azimuth}{\phi}                                    			
\newcommand{\azimuthAoA}{\varphi}                                  	
\newcommand{\azxAoA}[1]{\azimuthAoA_{ss',n}^{(#1)}}       	
\newcommand{\distVec}[0]{\V{d}_{n}^{(j)}}								
\newcommand{\aoaVec}[0]{\V{\varphi}_{n}^{(j)}}						
\newcommand{\aodVec}[0]{\V{\phi}_{n}^{(j)}}							
\newcommand{\varianceDist}{\sigma_{\mathrm{d}}^2} 			
\newcommand{\varianceAoA}{\sigma_{\mathrm{\varphi}}^2} 	
\newcommand{\varianceAoD}{\sigma_{\mathrm{\phi}}^2} 	    

\renewcommand{\observationMatrix}[1]{\V{z}_{#1}}				
\newcommand{\RVobservationMatrix}[1]{\RV{z}_{#1}}				
\renewcommand{\observation}[2]{\V{z}_{#1}^{(#2)}}   			

\renewcommand{\jacobOsubTX}[0]{ \jacobian_{\scriptscriptstyle n,j,\ncomponent}^{\scriptscriptstyle \text{\tiny o,TX}}}   

\newcommand{\pdfGaussian}{f_{\mathrm{N}}}					

\newcommand{\gainmatrix}[0]{\mathbf{\Gamma}}

\newcommand{\RVG}[1]{\mathrm{\boldsymbol{#1}}}	

\renewcommand{\etaglobal}[1]{\V{\theta}^{\text{\tiny{g}}}_{\scriptscriptstyle#1}}            
\renewcommand{\RVetaglobal}[1]{\RVG{\uptheta}^{\text{\tiny{g}}}_{\scriptscriptstyle#1}}            

\renewcommand{\etach}[2]{\V{\theta}^{\text{\tiny{ch}}}_{\scriptscriptstyle#1,#2}}            					
\newcommand{\RVetach}[2]{\RVG{\uptheta}^{\text{\tiny{ch}}}_{\scriptscriptstyle#1,#2}}           

\newcommand{\RVetaglobalHat}[1]{\hat{\RVG{\uptheta}}^{\text{\tiny{g}}}_{\scriptscriptstyle#1}}            

\newcommand{\RVstate}[1]{\RV{x}_{\scriptscriptstyle #1}}                                  
\renewcommand{\state}[1]{\V{x}_{\scriptscriptstyle #1}}                                  
\newcommand{\stateHat}[1]{\hat{\V{x}}_{\scriptscriptstyle #1}}                                  

\newcommand{\RVpMVAposStacked}[0]{\overline{\RV{p}}_{\mathrm{\surface}}}                   

\renewcommand{\processNoise}[1]{{\V{w}_{\scriptscriptstyle #1}}}
\newcommand{\RVprocessNoise}[1]{{\RV{w}_{\scriptscriptstyle #1}}}
\renewcommand{\state}[1]{\V{x}_{\scriptscriptstyle #1}}                                  

\newcommand{\noiseVarianceV}[0]{\mathrm{\sigma}_{\mathrm{v}}^2}				
\newcommand{\noiseVarianceO}[0]{\mathrm{\sigma}_{\mathrm{\scriptscriptstyle \Delta\varphi}}^{2}}				
\newcommand{\noiseVarianceP}[0]{\mathrm{\sigma}_{\mathrm{p}}^2}				

\renewcommand{\jacobO}[1]{ \jacobian_{\scriptscriptstyle n,j}^{{\text{\tiny o}},\scriptscriptstyle#1}}             
\renewcommand{\jacobP}[1]{ \jacobian_{\scriptscriptstyle n,j}^{{\text{\tiny p}},\scriptscriptstyle#1}}             
\renewcommand{\jacobM}[1]{ \jacobian_{\scriptscriptstyle n,j}^{{\text{\tiny m}},\scriptscriptstyle#1}}             
\newcommand{\interval}[1]{\hat{I}_{\scriptscriptstyle#1}}				

\usepackage{amsthm}

\newlength\figureheight
\newlength\figurewidth 

\newcommand{\gls}[1]{\ac{#1}}				
\newcommand{\glspl}[1]{\acp{#1}}			

\newlength{\plotWidth}

\newcommand{\circleblue}[1]{%
  \tikzexternaldisable%
  \tikz[baseline=(char.base)]{%
    \node[shape=circle,
          inner sep=0.5pt,
          minimum size=1.0em,
          fill=IEEEblue,
          text=white] (char) {\adjustbox{max width=0.65em}{#1}};
  }%
  \tikzexternalenable%
}

\newcounter{assumptions}
\newcommand{\nextassumption}[1][]{%
  \refstepcounter{assumptions}
  \circleblue{\arabic{assumptions}}~
  \if\relax\detokenize{#1}\relax%
  \else%
    \label{#1}
  \fi
}

\newcommand{\assref}[1]{\circleblue{\ref{#1}}}

\newcommand{\wallrefA}[0]{%
  \tikzexternaldisable%
  \tikz[baseline=(char.base)]{%
    \node[rectangle,inner sep=0.5pt,rounded corners=0.65mm,minimum size=1.0em,
          fill=IEEEblue,
          text=white] (char) {\adjustbox{max width=0.65em}{$1$}};
  }%
  \tikzexternalenable%
}

\newcommand{\wallrefC}[0]{%
  \tikzexternaldisable%
  \tikz[baseline=(char.base)]{%
    \node[rectangle,inner sep=0.5pt,rounded corners=0.65mm,minimum size=1.0em,
          fill=IEEEgreen,
          text=white] (char) {\adjustbox{max width=0.65em}{$3$}};
  }%
  \tikzexternalenable%
}

\newcommand{\wallrefE}[0]{%
  \tikzexternaldisable%
  \tikz[baseline=(char.base)]{%
    \node[rectangle,inner sep=0.5pt,rounded corners=0.65mm,minimum size=1.0em,
          fill=IEEEyellow,
          text=white] (char) {\adjustbox{max width=0.65em}{$5$}};
  }%
  \tikzexternalenable%
}

\definecolor{RDlightgreen}{RGB}{141 192 69}
\definecolor{RDgreen}{rgb}{0.3647, 0.4275, 0.2667}
\definecolor{RDdarkgreen}{rgb}{0.2196, 0.2196, 0.2196}
\definecolor{RDmaroon}{rgb}{.522,.22,.353} %

\definecolor{IEEEblue}{RGB}{0 98 155}
\definecolor{IEEElightblue}{RGB}{0 181 226}
\definecolor{IEEEturquoise}{RGB}{0 156 166}
\definecolor{IEEEred}{RGB}{186 12 47}
\definecolor{IEEEgreen}{RGB}{0 132 61}
\definecolor{IEEElightgreen}{RGB}{120 190 32}
\definecolor{IEEEorange}{RGB}{225 163 0}
\definecolor{IEEEyellow}{RGB}{255 209 0}
\definecolor{IEEEviolett}{RGB}{152 29 151}
\definecolor{IEEEdarkmaroon}{RGB}{134 31 65}
\definecolor{IEEEdarkorange}{RGB}{232 119 34}

\colorlet{red}{IEEEred}
\colorlet{blue}{IEEEblue}
\colorlet{orange}{IEEEorange}
\colorlet{green}{IEEEgreen}

\definecolor{sand}{rgb}{0.88235,0.63922,0.00000}
\colorlet{IEEEyellow}{sand}

\def\BibTeX{{\rm B\kern-.05em{\sc i\kern-.025em b}\kern-.08em
    T\kern-.1667em\lower.7ex\hbox{E}\kern-.125emX}}
\begin{document}


\title{
Posterior Cram\'er--Rao Bounds on Localization and Mapping Errors in Distributed MIMO SLAM \\ 
\thanks{
\IEEEauthorrefmark{1}Benjamin J. B. Deutschmann and Xuhong Li contributed equally to this work.
The AMBIENT-6G project has received funding from the Smart Networks and Services Joint Undertaking (SNS JU) under the European Union's Horizon Europe research and innovation programme under Grant Agreement No. 101192113.
This work was supported in part by the Knut and Alice Wallenberg Foundation.}
}

\author{\centering
\IEEEauthorblockN{1\textsuperscript{st} Benjamin J. B. Deutschmann\IEEEauthorrefmark{1}}
\IEEEauthorblockA{\textit{Institute of Comm. Networks and Satellite Comms.} \\
\textit{Graz University of Technology}\\
 Graz, Austria\\
benjamin.deutschmann@tugraz.at\,\orcidlink{0000-0002-2647-7662}}
\and
\IEEEauthorblockN{2\textsuperscript{nd} Xuhong Li\IEEEauthorrefmark{1}}
\IEEEauthorblockA{\textit{Department of Electrical and Computer Engineering} \\
\textit{University of California San Diego}\\
San Diego, USA \\
xul046@ucsd.edu\,\orcidlink{0000-0002-3095-7335}}
\and
\IEEEauthorblockN{3\textsuperscript{rd} Florian Meyer}
\IEEEauthorblockA{\textit{Department of Electrical and Computer Engineering} \\
\textit{University of California San Diego}\\
San Diego, USA \\
flmeyer@ucsd.edu\,\orcidlink{0000-0001-6985-2250}}
\and
\IEEEauthorblockN{4\textsuperscript{th} Erik Leitinger}
\IEEEauthorblockA{\textit{Institute of Comm. Networks and Satellite Comms.} \\
\textit{Graz University of Technology}\\
 Graz, Austria\\
erik.leitinger@tugraz.at\,\orcidlink{0000-0003-1048-4849}}
}

\allowdisplaybreaks
\frenchspacing
	\author{\IEEEauthorblockN{Benjamin J. B. Deutschmann\IEEEauthorrefmark{2}\IEEEauthorrefmark{1}, Xuhong Li\IEEEauthorrefmark{3}\IEEEauthorrefmark{1}, Florian Meyer\IEEEauthorrefmark{3}, Erik Leitinger$^\dagger$}
	\IEEEauthorblockA{
	\IEEEauthorrefmark{2}Institute of Comm. Networks and Satellite Comms.
	, Graz University of Technology, Austria \\ 
	\IEEEauthorrefmark{3}Department of Electrical and Computer Engineering, University of California San Diego, USA \\
	Email: benjamin.deutschmann@tugraz.at\,\orcidlink{0000-0002-2647-7662}, xul046@ucsd.edu\,\orcidlink{0000-0002-3095-7335}, flmeyer@ucsd.edu\,\orcidlink{0000-0001-6985-2250}, erik.leitinger@tugraz.at\,\orcidlink{0000-0003-1048-4849} }
	\vspace*{-6mm}}

\maketitle

\begin{abstract}
Radio-frequency simultaneous localization and mapping (RF-SLAM) 
methods jointly infer the position of mobile transmitters and receivers in wireless networks, together with a geometric map of the propagation environment.
An inferred map of specular surfaces can be used to exploit non-line-of-sight components of the multipath channel to increase robustness, bypass obstructions, and improve 
overall communication and positioning performance.
While performance bounds for user location are well established, the literature lacks performance bounds for map information.
This paper derives the mapping error bound (MEB), i.e., the posterior Cram\'er--Rao lower bound on the position and orientation of specular surfaces, for RF-SLAM. 
In particular, we consider a very general scenario with single- and double-bounce reflections, as well as distributed anchors.
We demonstrate numerically that a state-of-the-art RF-SLAM algorithm asymptotically converges to 
this MEB.
The bounds assess not only the localization (position and orientation) but also the mapping performance of RF-SLAM algorithms in terms of global features. 
\end{abstract}


\vspace*{-0.5mm}
\section{Introduction}\label{sec:intro}	

\subfile{./InputFiles/Introduction}

	\section{Geometrical Model of the Environment}
	\label{sec:GeometricalRelations}

\subfile{./InputFiles/GeometricalRelations}

\section{System model}	 
	\label{sec:SystemModel}%
	\subfile{./InputFiles/SystemModel}
\section{Posterior Cram\'er--Rao Lower Bound}	 
	\label{sec:PCRLB}
	\subfile{./InputFiles/PCRLB}

\section{Problem Formulation and Proposed Method}\label{sec:method}  

\subfile{./InputFiles/ProblemFormulation}

\section{Experiment and Results}\label{sec:results}  
	\subfile{./InputFiles/Results}

\section{Conclusion}\label{sec:conclusions}  
The main contribution of this paper is the derivation of a novel mapping error bound
, i.e., a lower bound on the estimation error for inferring the environment map in 
\ac{mpslam}.
Our efficient geometric surface model uses \acp{surface} 
which are global point map features. 
Unlike \acp{pcrlb} formulated for \acp{va}, our \ac{meb} defined in terms of these global features lets discriminate the mapping accuracy for individual specular surfaces even for double 
bounce reflections.
We have shown that our state-of-the-art SLAM algorithm is capable of asymptotically approaching the \ac{meb} on simulated data.

While the derivation of our \ac{meb} is analogously valid for \ac{3d} scenarios, future work will augment the \ac{oeb} for arbitrary agent rotations in \ac{3d} space.
In addition to the azimuth angles, \acp{aod} and \ac{aoa} will be extended by elevation angles.
On the methodological side, future work will augment our algorithm to capture wall extents.

\begin{appendices}
	\input{./InputFiles/appendix}

\end{appendices}

\bibliographystyle{IEEEtran}
\balance
\bibliography{IEEEabrv,references,referencesAsilomar}

\end{document}

%% file: definitions.tex
\usepackage{tikz}
\usetikzlibrary{backgrounds}
\tikzstyle{load}   = [ultra thick,-latex]
\tikzstyle{stress} = [-latex]
\tikzstyle{dim}    = [latex-latex]
\tikzstyle{axis}   = [-latex,black!55]

\definecolor{green(pigment)}{rgb}{0.0, 0.65, 0.31}
\definecolor{frenchblue}{rgb}{0.0, 0.45, 0.73} 
\definecolor{mediumcandyapplered}{rgb}{0.89, 0.02, 0.17}
\definecolor{darkGreyGreen598081}{rgb}{0.349, 0.502, 0.506}
\definecolor{darkGreyGreen2b6a99}{rgb}{0.349, 0.502, 0.506}
\definecolor{darkBlue1b7c3d}{rgb}{0.14, 0.35, 0.51}

\definecolor{redb11927}{rgb}{0.734, 0.104, 0.162}
\definecolor{blue4695be}{rgb}{0.171, 0.364, 0.465}

\usepackage{tkz-euclide}
\usetikzlibrary{arrows.meta}
\usepackage{setspace}
\usepackage{bm}
\usepackage{mathtools}
\usepackage{graphicx}
\usepackage{relsize}
\usepackage{cite}
\usepackage{amsmath}
\usepackage{color}
\usepackage{psfrag}
\usepackage{amssymb}
\usepackage{dblfloatfix}
\usepackage[font=small]{caption}
\usepackage{tabularx}
\usepackage{makecell}
\usepackage{lipsum}
\usepackage{float}
\usepackage{subfiles}
\usepackage{epstopdf}
\usepackage{verbatim}
\usepackage{amsmath}
\usepackage{ifthen}
\usepackage{pgfplots}
\usepackage{pgfplotstable}
\usepackage{tikzscale}
\usepackage{acronym}
\usepackage{amsfonts}
\usepackage{tikz-3dplot}
\usepackage{cuted}
\usepackage{subfig}

\usetikzlibrary{intersections}   
\usepgfplotslibrary{fillbetween}

\usetikzlibrary{spy,backgrounds}
\usetikzlibrary{fit}
\usetikzlibrary{calc}

\usetikzlibrary{3d}
\usetikzlibrary{shapes}
\usetikzlibrary {patterns,patterns.meta}
\usepackage{pgffor}
\usetikzlibrary{shapes.geometric}
\usetikzlibrary{angles,quotes}

\renewcommand{\baselinestretch}{1} 

\colorlet{veccol}{green!50!black}
\colorlet{projcol}{blue!70!black}
\colorlet{myblue}{blue!80!black}
\colorlet{myred}{red!90!black}
\colorlet{mydarkblue}{blue!50!black}
\tikzset{>=latex} 
\tikzstyle{proj}=[projcol!80,line width=0.08] 
\tikzstyle{area}=[draw=veccol,fill=veccol!80,fill opacity=0.6]
\tikzstyle{vector}=[-stealth,myblue,thick,line cap=round]
\tikzstyle{unit vector}=[->,veccol,thick,line cap=round]
\tikzstyle{dark unit vector}=[unit vector,veccol!70!black]
\usetikzlibrary{angles,quotes} 

\definecolor{RDlightgreen}{RGB}{141 192 69}
\definecolor{RDgreen}{rgb}{0.3647, 0.4275, 0.2667}
\definecolor{RDdarkgreen}{rgb}{0.2196, 0.2196, 0.2196}
\definecolor{RDmaroon}{rgb}{.522,.22,.353} %

\definecolor{IEEEblue}{RGB}{0 98 155}
\definecolor{IEEElightblue}{RGB}{0 181 226}
\definecolor{IEEEturquoise}{RGB}{0 156 166}
\definecolor{IEEEred}{RGB}{186 12 47}
\definecolor{IEEEgreen}{RGB}{0 132 61}
\definecolor{IEEElightgreen}{RGB}{120 190 32}
\definecolor{IEEEorange}{RGB}{225 163 0}
\definecolor{IEEEyellow}{RGB}{255 209 0}
\definecolor{IEEEviolett}{RGB}{152 29 151}
\definecolor{IEEEdarkmaroon}{RGB}{134 31 65}
\definecolor{IEEEdarkorange}{RGB}{232 119 34}

\newcommand{\ist}{\hspace*{.3mm}}
\newcommand{\rmv}{\hspace*{-.3mm}}
\newcommand{\iist}{\hspace*{1mm}}

\newcommand{\nn}{\nonumber}

\DeclareMathAlphabet{\mathsfbr}{OT1}{cmss}{m}{n}
\SetMathAlphabet{\mathsfbr}{bold}{OT1}{cmss}{bx}{n}
\DeclareRobustCommand{\msf}[1]{%
	\ifcat\noexpand#1\relax\msfgreek{#1}\else\mathsfbr{#1}\fi
}

\makeatletter
\newcommand{\msfgreek}[1]{\csname s\expandafter\@gobble\string#1\endcsname}
\makeatother

\DeclareRobustCommand{\mcal}[1]{%
	\ifcat\noexpand#1\relax\mathnormal{#1}\else\cal{#1}\fi
}
\DeclareRobustCommand{\BM}[1]{%
	\ifcat\noexpand#1\relax\bm{\boldUppercaseItalicGreek{#1}}\else\bm{#1}\fi
}
\makeatletter
\newcommand{\boldUppercaseItalicGreek}[1]{\csname var\expandafter\@gobble\string#1\endcsname}
\makeatother

\newcommand{\rv}[1]{\msf{#1}}
\newcommand{\RV}[1]{\bm{\msf{#1}}}

\newcommand{\V}[1]{\bm{#1}}
\newcommand{\M}[1]{\BM{#1}}

\let\geq\geqslant

\let\leq\leqslant

\let\succeq\succcurlyeq

\providecommand{\IEEEQED}{\IEEEQEDopen}

\DeclareMathVersion{timesmath}
\SetSymbolFont{letters}{timesmath}{OML}{ntxmi}{m}{it}
\DeclareMathVersion{timesmathbold}
\SetSymbolFont{letters}{timesmathbold}{OML}{ntxmi}{b}{it}

\makeatletter
\newif\ifAC@uppercase@first%
\def\Aclp#1{\AC@uppercase@firsttrue\aclp{#1}\AC@uppercase@firstfalse}%
\def\AC@aclp#1{%
	\ifcsname fn@#1@PL\endcsname%
	\ifAC@uppercase@first%
	\expandafter\expandafter\expandafter\MakeUppercase\csname fn@#1@PL\endcsname%
	\else%
	\csname fn@#1@PL\endcsname%
	\fi%
	\else%
	\AC@acl{#1}s%
	\fi%
}%
\def\Acp#1{\AC@uppercase@firsttrue\acp{#1}\AC@uppercase@firstfalse}%
\def\AC@acp#1{%
	\ifcsname fn@#1@PL\endcsname%
	\ifAC@uppercase@first%
	\expandafter\expandafter\expandafter\MakeUppercase\csname fn@#1@PL\endcsname%
	\else%
	\csname fn@#1@PL\endcsname%
	\fi%
	\else%
	\AC@ac{#1}s%
	\fi%
}%
\def\Acfp#1{\AC@uppercase@firsttrue\acfp{#1}\AC@uppercase@firstfalse}%
\def\AC@acfp#1{%
	\ifcsname fn@#1@PL\endcsname%
	\ifAC@uppercase@first%
	\expandafter\expandafter\expandafter\MakeUppercase\csname fn@#1@PL\endcsname%
	\else%
	\csname fn@#1@PL\endcsname%
	\fi%
	\else%
	\AC@acf{#1}s%
	\fi%
}%
\def\Acsp#1{\AC@uppercase@firsttrue\acsp{#1}\AC@uppercase@firstfalse}%
\def\AC@acsp#1{%
	\ifcsname fn@#1@PL\endcsname%
	\ifAC@uppercase@first%
	\expandafter\expandafter\expandafter\MakeUppercase\csname fn@#1@PL\endcsname%
	\else%
	\csname fn@#1@PL\endcsname%
	\fi%
	\else%
	\AC@acs{#1}s%
	\fi%
}%
\edef\AC@uppercase@write{\string\ifAC@uppercase@first\string\expandafter\string\MakeUppercase\string\fi\space}%
\def\AC@acrodef#1[#2]#3{%
	\@bsphack%
	\protected@write\@auxout{}{%
		\string\newacro{#1}[#2]{\AC@uppercase@write #3}%
	}\@esphack%
}%
\def\Acl#1{\AC@uppercase@firsttrue\acl{#1}\AC@uppercase@firstfalse}
\def\Acf#1{\AC@uppercase@firsttrue\acf{#1}\AC@uppercase@firstfalse}
\def\Ac#1{\AC@uppercase@firsttrue\ac{#1}\AC@uppercase@firstfalse}
\def\Acs#1{\AC@uppercase@firsttrue\acs{#1}\AC@uppercase@firstfalse}
\robustify\Aclp
\robustify\Acfp
\robustify\Acp
\robustify\Acsp
\robustify\Acl
\robustify\Acf
\robustify\Ac
\robustify\Acs
\makeatother

\makeatletter
\def\underbracex#1#2{\mathop{\vtop{\m@th\ialign{##\crcr
				$\hfil\displaystyle{#2}\hfil$\crcr
				\noalign{\kern3\p@\nointerlineskip}%
				#1\crcr\noalign{\kern3\p@}}}}\limits}

\def\upbracefilla{$\m@th \setbox\z@\hbox{$\braceld$}%
	\bracelu\leaders\vrule \@height\ht\z@ \@depth\z@\hfill 
	\leaders\vrule \@height\ht\z@ \@depth\z@\hfill\bracerd
	\braceld\leaders\vrule \@height\ht\z@ \@depth\z@\hfill
	\kern\p@\vrule \@width\p@\kern\p@\vrule \@width\p@\kern\p@\vrule \@width\p@ 
	$}

\def\upbracefilld{$\m@th \setbox\z@\hbox{$\braceld$}%
	\vrule \@width\p@\kern\p@\vrule \@width\p@\kern\p@\vrule \@width\p@\kern\p@
	\leaders\vrule \@height\ht\z@ \@depth\z@\hfill\braceru$}

\makeatother

\acrodef{2d}[2D]{two-dimensional}
\acrodef{3d}[3D]{three-dimensional}
\acrodef{5g}[5G]{5th-generation}
\acrodef{pnt}[PNT]{positioning, navigation and timing}
\acrodef{pa}[PA]{physical anchor}
\acrodef{bs}[BS]{base station}
\acrodef{awgn}[AWGN]{additive white Gaussian noise}
\acrodef{va}[VA]{virtual anchor}
\acrodef{mpc}[MPC]{multipath component}
\acrodef{fa}[FA]{false alarm}
\acrodef{far}[FAR]{false alarm rate}
\acrodef{pva}[PR]{potential ray}
\acrodef{mf}[MF]{map feature}
\acrodef{pmf}[PMF]{potential \ac{mf}}
\acrodef{smc}[SMC]{specular multipath component}
\acrodef{psmc}[PSMC]{potential \ac{smc}}
\acrodef{mva}[SFV]{surface feature vector}
\acrodef{ip}[IP]{interaction point}
\acrodef{pmva}[PSFV]{potential \ac{mva}}
\acrodef{slam}[SLAM]{simultaneous localization and mapping}
\acrodef{Mp}[MP]{Multipath-based}
\acrodef{mpslam}[MP-SLAM]{multipath-based simultaneous localization and mapping}
\acrodef{pmf}[PMF]{probability mass function}
\acrodef{pdf}[PDF]{probability density function}
\acrodef{cdf}[CDF]{cummulative distribution function}
\acrodef{bp}[BP]{belief propagation}
\acrodef{spa}[SPA]{sum-product algorithm}
\acrodef{fg}[FG]{factor graph}
\acrodef{mmse}[MMSE]{minimum mean-square error}
\acrodef{simo}[SIMO]{single-input-multiple-output}
\acrodef{miso}[MISO]{multiple-input-single-output}
\acrodef{mimo}[MIMO]{multiple-input-multiple-output}
\acrodef{dmimo}[D-MIMO]{distributed multiple-input-multiple-output}
\acrodef{mmwave}[mmWave]{millimeter-wave}
\acrodef{ue}[UE]{mobile user}
\acrodef{los}[LoS]{line-of-sight}
\acrodef{nlos}[NLoS]{non \ac{los}}
\acrodef{aoa}[AoA]{angle-of-arrival}
\acrodef{aod}[AoD]{angle-of-departure}
\acrodef{roi}[RoI]{region-of-interest}
\acrodef{snr}[SNR]{signal-to-noise ratio}
\acrodef{ospa}[OSPA]{optimal subpattern assignment}
\acrodef{mospa}[MOSPA]{mean \ac{ospa}}
\acrodef{dmc}[DMC]{dense multipath component}
\acrodef{sr}[SR]{super-resolution}
\acrodef{sbl}[SBL]{sparse Bayesian learning}
\acrodef{sde}[SDE]{sequential detection and estimation}
\acrodef{kest}[KEST]{Kalman enhanced super resolution tracking}
\acrodef{da}[DA]{data association}
\acrodef{fim}[FIM]{Fisher information matrix}
\acrodef{crlb}[CRLB]{Cram\'{e}r–-Rao lower bound}
\acrodef{pcrlb}[PCRLB]{posterior Cram\'{e}r–-Rao lower bound}
\acrodef{meb}[MEB]{mapping error bound}
\acrodef{peb}[PEB]{position error bound}
\acrodef{oeb}[OEB]{orientation error bound}
\acrodef{va}[VA]{virtual anchor}
\acrodef{eadf}[EADF]{effective aperture distribution function}
\acrodef{lhf}[LHF]{likelihood function}
\acrodef{rmse}[RMSE]{root mean square error} 
\acrodef{mse}[MSE]{mean square error} 
\acrodef{ps}[PS]{point scatterer}
\acrodef{imm}[IMM]{interacting multiple model}
\acrodef{rf}[RF]{radio frequency}
\acrodef{iid}[iid]{independent and identically distributed}
\acrodef{prop}[PROP]{proposed algorithm}
\acrodef{refa}[REFA]{reference algorithm}
\acrodef{rt}[RT]{ray tracing}
\acrodef{ceda}[CEDA]{snapshot-based parametric channel estimation and detection algorithm}

%% file: InputFiles/math-notation.tex
\newcommand{\PCRLB}[0]{\CRLB_{\scriptscriptstyle n|n}}                 
\newcommand{\FIMstep}[2]{\bm{\mathcal{I}}_{\scriptscriptstyle #1|#2}}     
\newcommand{\rxTX}[0]{\grave{{r}}_{\scriptscriptstyle x}}
\newcommand{\ryTX}[0]{\grave{{r}}_{\scriptscriptstyle y}}

\newcommand{\jacobO}[1]{ \bm{O}_{\scriptscriptstyle n,j}^{{\text{\tiny TX}}\scriptscriptstyle#1}}             
\newcommand{\OEBRX}[0]{\sigma_o}                   

\newcommand{\VEB}[0]{\sigma_v}                                      			 	

\newcommand{\jacobOsubTX}[0]{ \jacobian_{\scriptscriptstyle n,j,\MVApair{s}{s'}}^{\scriptscriptstyle \text{\tiny o,TX}}}   

\newcommand{\rotMdot}[1]{\dot{\bm{M}}_{\scriptscriptstyle #1}}                                       

\newcommand{\eye}[1]{\mathbf{I}_{\scriptscriptstyle #1}}                
\newcommand{\diff}[0]{\mathrm{d}}       

\newcommand{\lightspeed}{\mathsf{c}}                                

\newcommand{\dimLocal}[0]{ \mathrm{D}_\text{\tiny{ch}} }            
\newcommand{\dimGlobal}[0]{ \mathrm{D}_\text{\tiny{g}} }            
\newcommand{\Ncomponents}[0]{ K }                          
\newcommand{\ncomponent}[0]{ \kappa }                               

\newcommand{\setWalls}[0]{ \mathcal{S} }                            
\newcommand{\setSDB}[0]{ \mathcal{D} }                              
\newcommand{\setSDBt}[0]{ \widetilde{\mathcal{D}} }                 
\newcommand{\setSB}[0]{ \mathcal{D}_{\text{S}} }                    
\newcommand{\setDB}[0]{ \mathcal{D}_{\text{D}} }                    

\newcommand{\MVApair}[2]{(\!#1,#2\!)}                               

\newcommand{\pmva}[1]{\bm{p}^{\text{\tiny mva}}_{\scriptscriptstyle #1}}               
\newcommand{\hva}[0]{{h}_{\text{\tiny va}}}                         

\newcommand{\origin}[0]{\bm{0}}                                     
\newcommand{\rotM}[0]{\bm{M}_{\scriptscriptstyle j}}                                       

\newcommand{\posAnchor}[3]{ {\bm{p}_{\text{\tiny c}\scriptscriptstyle,#1}^{\scriptscriptstyle\MVApair{#2}{#3}}} }     
\newcommand{\range}[4]{\bm{r}_{\scriptscriptstyle#1,#2}^{\scriptscriptstyle\MVApair{#3}{#4}}} 
\newcommand{\ranget}[4]{\widetilde{\bm{r}}_{\scriptscriptstyle#1,#2}^{\scriptscriptstyle\MVApair{#3}{#4}}} 
\newcommand{\rangepTX}[4]{\grave{\bm{r}}_{\scriptscriptstyle#1,#2}^{\scriptscriptstyle\MVApair{#3}{#4}}} 
\newcommand{\rangep}[4]{\acute{\bm{r}}_{\scriptscriptstyle#1,#2}^{\scriptscriptstyle\MVApair{#3}{#4}}} 
\newcommand{\rx}[0]{\acute{{r}}_{\scriptscriptstyle x}}
\newcommand{\ry}[0]{\acute{{r}}_{\scriptscriptstyle y}}

\newcommand{\er}[4]{\bm{e}_{\scriptscriptstyle#1,#2}^{\scriptscriptstyle\MVApair{#3}{#4}}} 
\newcommand{\ert}[4]{\widetilde{\bm{e}}_{\scriptscriptstyle#1,#2}^{\scriptscriptstyle\MVApair{#3}{#4}}} 
\newcommand{\house}[1]{ \bm{\mathcal{H}}_{\scriptscriptstyle #1} }   

\newcommand{\observation}[2]{\bm{z}_{\scriptscriptstyle #1}^{\scriptscriptstyle(#2)}}   
\newcommand{\observationMatrix}[1]{\bm{Z}_{\scriptscriptstyle #1}}                      

\newcommand{\PMVAposStacked}[1]{\overline{\bm{p}}^{\text{\tiny mva}}_{\scriptscriptstyle \,}}                   
\newcommand{\state}[1]{\bm{x}_{\scriptscriptstyle #1}}                                  
\newcommand{\pos}[1]{ \bm{p}_{\scriptscriptstyle #1} }                                  
\newcommand{\vel}[1]{ \bm{v}_{\scriptscriptstyle #1} }                                  

\newcommand{\etaglobal}[1]{\parvec^{\text{\tiny{g}}}_{\scriptscriptstyle#1}}            
\newcommand{\RVetaglobal}[1]{\RV{\parvec}^{\text{\tiny{g}}}_{\scriptscriptstyle#1}}            
\newcommand{\etach}[2]{\parvec^{\text{\tiny{ch}}}_{\scriptscriptstyle#1,#2}}            

\newcommand{\CRLB}[0]{\bm{P}}                      
\newcommand{\FIMch}[2]{\bm{\mathcal{I}}_{\scriptscriptstyle #1,#2}^{\text{\tiny ch}}}   
\newcommand{\FIMglobal}[1]{\bm{\mathcal{I}}_{\scriptscriptstyle #1}^{\text{\tiny g}}}   

\newcommand{\jacobian}{ \boldsymbol{J} }
\newcommand{\jacobgn}[1]{ \jacobian_{\scriptscriptstyle n,\scriptscriptstyle#1}}             
\newcommand{\jacobP}[1]{ \bm{P}_{\scriptscriptstyle n,j}^{\scriptscriptstyle#1}}             
\newcommand{\jacobM}[1]{ \bm{M}_{\scriptscriptstyle n,j}^{\scriptscriptstyle#1}}             
\newcommand{\jacobMVAblock}[2]{ \jacobian_{\scriptscriptstyle n,j}^{\scriptscriptstyle\MVApair{#1}{#2}}}   
\newcommand{\jacobMVAblockG}[2]{ \V{G}_{\scriptscriptstyle n,j}^{\scriptscriptstyle\MVApair{#1}{#2}}}   
\newcommand{\jacobMVAsb}[0]{ \jacobian_{\scriptscriptstyle n,j}^{\scriptscriptstyle \text{\tiny sb},s}}   

\newcommand{\PEB}[0]{\sigma_p}                                      
\newcommand{\MEB}[1]{\sigma_{\bm{p}_s}^{\scriptscriptstyle(#1)}}          

\newcommand{\azimuth}{\vartheta}                                    

\newcommand{\azx}[1]{\azimuth_{\scriptscriptstyle n,#1}^{\MVApair{s}{s'}}}               
\newcommand{\delayx}[1]{\widetilde{\tau}_{\scriptscriptstyle n,#1}^{\MVApair{s}{s'}}}    

\newcommand{\delay}[0]{\tau}        

\newcommand{\pwk}[0]{ \bm{p}^\text{\tiny{w}}_{\scriptscriptstyle s} }                  
\newcommand{\nw}[0]{ {\bm{n}^\text{\tiny{w}}_{\scriptscriptstyle s}} }                   
\newcommand{\nwc}[1]{ {\bm{n}^\text{\tiny{w}}_{\scriptscriptstyle {#1}}} }                   

\newcommand{\transitionmatrix}{{\mathbf{\Phi}}}                         
\newcommand{\processNoiseCov}{{\mathbf{Q}}}                             
\newcommand{\processNoise}[1]{{\bm{\omega}_{\scriptscriptstyle #1}}}



%% file: InputFiles/Introduction.tex
Mapping---inferring a representation of the environment---
represents a viable component in providing location-aware information services to mobile users, enabling a range of use-case-specific applications~\cite{Amjad23RadioSLAMreview}. 
Beyond application-driven objectives, obtaining an environment map also serves a more fundamental purpose in wireless communications by admitting a geometric representation of the multipath channel. 
Particularly in indoor scenarios, measurements show that the dominant fraction of signal power transmitted over a multipath channel resides in the \ac{los} and a few specular \ac{nlos} components~\cite{Gentile18NLoS}. 
This makes mapping particularly vital in bypassing \ac{olos} conditions by exploiting \ac{nlos} components, providing robustness in localizing the agent (i.e., the mobile user equipment) in \ac{rf} \ac{mpslam}~\cite{Masiero25SPslam}.
Geometric map information also benefits physical-layer communications: optimal conjugate downlink beamforming~\cite{mMIMObook} jointly uses the \ac{los} and \ac{nlos} components~\cite{Deutschmann25WCM}. 
While conjugate beamforming using noisy channel observations (i.e., reciprocity beamforming) becomes power-optimal in static scenarios and under high \ac{snr}, it suffers from user mobility and low \ac{snr}.
\ac{mpslam} can become a key enabler 
in making predictions of the wireless multipath channel and providing mobility support, robustness, and high data rates even under harsh channel conditions~\cite{Palleit11Prediction,DeutschmannOJSP2025}.
Accurate mapping is essential for delivering these capabilities---and motivates the need for fundamental performance limits on map inference, allowing to benchmark \ac{mpslam} algorithms.

In most existing \ac{mpslam} frameworks~\cite{Ge22mmWaveSLAM,Masiero25SPslam,Erik_SLAM_TWC2019,XuhongTWC2025}, objects interacting with \ac{rf} signals are typically represented by {independent} point map features, with each feature producing a single \ac{rf} signal measurement. 
A commonly used map feature type is the \ac{va}, an image of a \ac{pa} mirrored across flat surfaces, which models specularly reflected \ac{rf} signals.
In the literature, there exist explicit performance bounds such as the \ac{crlb} or \ac{pcrlb} for the localization problem in \ac{mpslam}, i.e., the \ac{peb} and \ac{oeb} on the agent state~\cite{Guerra18OEB}. 
However, performance bounds for the mapping problem are usually formulated in terms of \acp{va}~\cite{GentnerTWC2016}, which do not capture how well each surface has been jointly inferred by multiple \acp{pa}, for two reasons: (i) \acp{va} are features that are \textit{local} to each \ac{pa} meaning that they prohibit a data fusion of the map-related information they encode across multiple \acp{pa}~\cite[Sec.\,VI\,B.]{LeiVenTeaMey:TSP2023}, which becomes particularly relevant in \ac{dmimo}. (ii) In general, each propagation path needs to be represented by a separate \ac{va}.
While single-bounce paths involve reflections at only a single surface, paths of double- and higher-order bounces involve reflections at \textit{multiple} surfaces with respective \acp{va} simultaneously ``bundling'' information about multiple map features.
However, to quantify how well each individual map feature can be inferred globally by all \acp{pa}, we are interested in performance bounds on the mapping error for (i) \textit{global} features, allowing data fusion across multiple \acp{pa} and propagation paths, where each feature represents only (ii) a \textit{single} surface, bounding the estimation performance of individual surfaces.

\pagebreak

{\slshape Contributions.} 
In this paper, we derive the \ac{pcrlb} for \ac{mpslam} using the \ac{surface} model, formerly termed \ac{mva} model, introduced in~\cite{LeiVenTeaMey:TSP2023}, which is a global point map feature representing only a single surface. With this model, we perform data fusion across \acp{mpc} of up to second-order bounces and across multiple distributed \acp{pa}.
The key contributions of this paper are as follows.
\begin{itemize}
	\item We derive the \ac{pcrlb} in terms of the \ac{peb} and \ac{oeb} on the agent state as well as the \ac{meb} 
	providing a novel lower bound on the \textit{mapping} error in terms of the \ac{surface} model from~\cite{LeiVenTeaMey:TSP2023}.
	\item Using synthetic data, we benchmark our \ac{mpslam} algorithm---described in \cite{XuhongTWC2025} and capable of fusing measurements from multiple \acp{pa} and propagation paths---against the derived \ac{pcrlb}.
	Our results show that this state-of-the-art \ac{mpslam} method 
is capable of approaching the \acp{pcrlb}.
\end{itemize}

\textit{Notations}: Column vectors and matrices are denoted as lowercase and uppercase bold letters. 
We use $\left[ \bm{X}\right]_{i,j}$ to denote the element of row $i$ and column $j$ in matrix $\bm{X}$. 
Random variables are displayed in san serif, upright fonts as for example $\rv{x}$ and $\RV{x}$ and their realizations in serif, italic font as for example $x$ and $\V{x}$. $f(\V{x})$ denotes the \ac{pdf} or \ac{pmf} of continuous or discrete random vector. $ \vert\cdot\vert $ represents the cardinality of a set. $f_{\mathrm{N}}(z|a,b)$ denotes Gaussian \ac{pdf} with mean $a$ and variance $b^2$.

%% file: InputFiles/GeometricalRelations.tex
We consider a \ac{2d} \ac{dmimo} \ac{dl} SLAM scenario. 
The distributed \acp{pa} $j \in \{1,\dots,J\}$ have known positions $\V{p}_{\mathrm{pa}}^{(j)} =[p_{\mathrm{pa},\mathrm{x}}^{(j)} \iist p_{\mathrm{pa},\mathrm{y}}^{(j)}]^{\mathrm{T}} \rmv\in\rmv \mathbb{R}^{2\rmv\times\rmv1}$ and known orientations $\Delta\phi^{(j)}$, and they are equipped with $ H_{\mathrm{tx}} $-element antenna arrays.
At each discrete time $n$, they transmit \ac{dl} radio signals which we assume to propagate in the horizontal $xy$-plane and impinge on the mobile agent.
The agent has 
unknown time-varying position $\V{p}_{n} =[p_{n,\mathrm{x}} \iist p_{n,\mathrm{y}}]^{\mathrm{T}} \rmv\in\rmv \mathbb{R}^{2\rmv\times\rmv1} $ and orientation $\Delta\varphi_{n}$, respectively.
It is equipped with a $ H_{\mathrm{rx}} $-element antenna array and acts as a receiver.
The positions $ \V{p}_{\mathrm{pa}}^{(j)} $ and $ \V{p}_{n} $ represent the phase-center positions of the respective arrays. 
We use $ \V{v}_{n}$ to denote the unknown time-varying agent velocity at time $n$ which evolves independently of the agent orientation $\Delta\varphi_{n}$. 
In this work, we assume perfect time and frequency synchronization between the clocks of all \acp{pa} and the agent, but 
no phase calibration is available, which only allows for a non-coherent data fusion among \acp{pa}~\cite{Fascista25RadioStripes}.

{\slshape Surface Model.}
We model a set $\setWalls\!:=\!\{1\dots S\}$ of specular surfaces $s\in\setWalls$ that give rise to single-bounce reflections $\setSB\!:=\!\{\MVApair{s}{s} \in \setWalls^2 \}$ and double-bounce\footnote{Propagation paths up to second-order reflections constitute the primary contribution to the received signals, making this a practical assumption. However, the model can be extended to incorporate higher-order bounce paths, though this introduces a substantial increase in complexity.} reflections $\setDB\!:=\!\{\MVApair{s}{s'} \in \setWalls^2 | s\neq s' \}$.
The set of all \acp{mpc} at the mobile agent from the \ac{dl} signal of anchor $j$ are $\setSDBt \!:=\! \{\MVApair{0}{0}\} \cup \setSDB$ with $\MVApair{0}{0}$ denoting the \ac{los} and $\setSDB\!:=\!\setSB \cup \setDB$.
Hence, the maximum number of \acp{mpc} 
is $\Ncomponents\!:=\!|\setSDBt| \! \triangleq \!  1+S^2$.
For brevity, we define $ \posAnchor{j} := \posVA{j}{0}{0}  $. 
From the perspective of the agent, signals are virtually impinging from the positions $\posVA{j}{s}{s}$ of single-bounce \acp{va} and the positions $\posVA{j}{s}{s'}$ of double-bounce \acp{va} which are images of \acp{pa}, virtually mirrored across surfaces $s$ and $s'$.

The \ac{surface} model (previously termed \ac{mva} model) from~\cite{LeiVenTeaMey:TSP2023} represents a specular surface $s$ solely through the \ac{surface} position $\pmva{s}$, which is computed by mirroring the origin $\origin$ of the global Cartesian coordinate system across surface $s$.
It elegantly represents both the position of the surface through a wall-point 
    $\pwk = \frac{\pmva{s}}{2}  \in \realset{2}{1}$
and the surface orientation\footnote{The surface orientation angle is $\vartheta_s := \arctantwo \big([\pmva{s}]_2,[\pmva{s}]_1\big).$} through a normal vector
    $\nw = \frac{\pmva{s}}{\lVert \pmva{s} \rVert}  \in  \realset{2}{1}$
with only a single variable $\pmva{s} \in \realset{2}{1}$~\cite{DeutschmannOJSP2025}.
The model is easily extended with two scalar lengths representing the surface extents to both sides of $\pwk$.

\subsubsection{Line-of-Sight Paths}\noindent
Following~\cite{DeutschmannCISA2025}, let 
$\range{n}{j}{0}{0}\!:=\!\pos{n}-\posAnchor{j}$ be the \ac{los} vector pointing from \ac{pa} $j$ to the agent at time $n$ in \textit{global} Cartesian coordinates, the vector $\rangep{n}{j}{0}{0} \!=\! \rotM^{-1} \range{n}{j}{0}{0}$ points from \ac{pa} $j$ to the agent in \textit{local} Cartesian coordinates (i.e., the local frame of reference) of \ac{pa} $j$.
We model the orientation of \ac{pa} $j$ in global coordinates using the rotation matrix $\rotM\!:=\!\bm{R}(\Delta\phi^{(j)})\!\in\!SO(2)$ defined in Appendix~\ref{eq:azRotM}, which is from the special orthogonal group $SO(2)\!=\!\{\bm{M}\!\in\!\mathbb{R}^{2 \times 2}|\bm{M} \bm{M}^\trp =\bm{M}^\trp \bm{M}\!=\!\eye{2}, \det(\bm{M})\!=\!1 \}$.
The respective \ac{aod} at transmitting \ac{pa} $j$ is
\begin{align}
\aod{0}{0} 
:= \arctantwo \big([\rangep{n}{j}{0}{0}]_2,[\rangep{n}{j}{0}{0}]_1\big) \,,
\end{align}
while $\dist{0}{0}\!:=\!\lVert \range{n}{j}{0}{0} \rVert$ is the \ac{los} distance to the \ac{pa}.
The \ac{aoa} at the receiving agent is
\begin{align}
	\aoa{0}{0} 
	:= \arctantwo \big([\rangepRX{n}{j}{0}{0}]_2,[\rangepRX{n}{j}{0}{0}]_1\big) 
\end{align}
with $\rangepRX{n}{j}{0}{0} \!=\! \rotMn^{-1} (-\range{n}{j}{0}{0})$ pointing from the receiving agent to the transmitting \ac{pa} $j$ in \textit{local} Cartesian agent coordinates.
We model the orientation of the agent at time $n$ in global coordinates using the rotation matrix $\rotMn\!:=\!\bm{R}(\Delta\varphi_{n})\!\in\!SO(2)$.  
The distance, \ac{aoa}, and \ac{aod} (i.e., $\{\dist{0}{0}, \aoa{0}{0}, \aod{0}{0} \}$ for the \ac{los} case) denote the \textit{local} channel parameters perceived on \ac{dl} observations. 

\subsubsection{Single-Bounce Paths}\noindent
Reflections of the \ac{dl} signals of \ac{pa} $j$ at large specular surfaces $s$ are modeled as if they were 
impinging at the agent from the positions of \acp{va} that are images of \acp{pa} mirrored across surfaces $s$.
We model \acp{va} using \acp{surface} as described in~\cite{LeiVenTeaMey:TSP2023},\cite{DeutschmannOJSP2025}:
The transformation from a \ac{pa} to a \ac{va} phase center position $\posVA{j}{s}{s} \!=\! \hva(\posAnchor{j},\pmva{s})$ is computed using the function $\hva \!:\! \mathbb{R}^2 \!\times\! \mathbb{R}^2 \!\to\! \mathbb{R}^2$
defined as
\begin{align}\label{eq:posAnchor-SB}
\hva(\posAnchor{j},\pmva{s})  \!=\! \posAnchor{j} \!-\!\left(\! 
        \frac{2  {\posAnchor{j}}^\trp \pmva{s}}{\lVert \pmva{s} \rVert^2} \!-\! 1\!
    \right)  \pmva{s} .
\end{align}
The Householder matrix
\begin{align}\label{eq:house}
    \house{s} = \eye{2} - 2 \, \frac{{\pmva{s}} {\pmva{s}}^\trp}{\lVert \pmva{s} \rVert^2}
\end{align}
represents the transformation from the \ac{pa} orientation to the \ac{va} orientation when mirrored across specular surface $s$.
For compatibility, let $\house{0}\!=\!\eye{2}$ for the \gls{los} path.
The distances, \acp{aoa}, and \acp{aod} of single-bounce paths are
\begin{align}
    \dist{s}{s}(\pos{n},\pmva{s}) &=\lVert \range{n}{j}{s}{s} \rVert  \, ,\\
	\aoa{s}{s}(\pos{n},\pmva{s}) &= \arctantwo \big([\rangepRX{n}{j}{s}{s}]_2,[\rangepRX{n}{j}{s}{s}]_1\big) \, ,\\
	\aod{s}{s}(\pos{n},\pmva{s}) &= \arctantwo \big([\rangep{n}{j}{s}{s}]_2,[\rangep{n}{j}{s}{s}]_1\big) \,,
\end{align}
respectively, with the spatial vectors (cf.\,Table~\ref{tab:spatial-vectors}) 
\begin{itemize}
\item $\range{n}{j}{s}{s}\!=\!\pos{n}-\posVA{j}{s}{s}$ pointing from the single-bounce \ac{va} to the ``physical agent'', abbreviated \ac{pm}, in \textit{global} Cartesian coordinates,
\item $\ranget{n}{j}{s}{s}\!=\house{s} \range{n}{j}{s}{s}$ pointing from \ac{pa} $j$ to the single-bounce ``virtual agent'', abbreviated \ac{vm}, in \textit{global} Cartesian coordinates,
\item $\rangep{n}{j}{s}{s}\!=\!\rotM^{-1} \big( \ranget{n}{j}{s}{s} \big)$ pointing from the \ac{pa} to the single-bounce \ac{vm} in \textit{local} Cartesian \ac{pa} coordinates, and
\item $\rangepRX{n}{j}{s}{s}\!=\! \rotMn^{-1} \big( -\range{n}{j}{s}{s} \big)$ pointing from the agent to the single-bounce \ac{va} in \textit{local} Cartesian agent coordinates.
\end{itemize}

\begin{table}
	\centering
    \caption{Spatial vectors in global or local Cartesian coordinates.}
	\label{tab:spatial-vectors}
    \begin{tabularx}{0.8\columnwidth}{ccc|ccc}
        \toprule 
        \multicolumn{3}{c|}{Global coordinates} & \multicolumn{3}{c}{Local coordinates}   \\
		Vector & from & to & Vector & from & to \\
    \midrule 
		$\range{n}{j}{s}{s'}$ 	& \ac{va} 	& \ac{pm} \\
		$-\range{n}{j}{s}{s'}$ & \ac{pm} & \ac{va} 	& $\rangepTX{n}{j}{s}{s'}$ & \ac{pm} 	& \ac{va} \\
		$\ranget{n}{j}{s}{s'}$ & \ac{pa} 	& \ac{vm} & $\rangep{n}{j}{s}{s'}$ 		& \ac{pa} 		& \ac{vm} \\
    \bottomrule
    \end{tabularx}
    \vspace{-5mm}
\end{table}

\begin{figure}[tb]
    \setlength{\plotWidth}{1.5\linewidth}
    \resizebox{\linewidth}{!}{\input{figures/SMC_incidence_angles}}
    \vspace{-0.5cm}\caption{One exemplary double-bounce path $\MVApair{s}{s'}$ with respective single-bounce and double-bounce \acp{va} and \acp{vm}.
    On the \ac{dl}, the spatial vector $-\range{n}{j}{s}{s'}$ maps to \ac{aoa} $\aoa{s}{s'}$ and the vector $\ranget{n}{j}{s}{s'}$ maps to \ac{aod} $\aod{s}{s'}$.
    }
     \label{fig:SMC_incidence_angles}
\end{figure}

\subsubsection{Double-Bounce Paths}\noindent

As described in~\cite{LeiVenTeaMey:TSP2023}, a double-bounce \ac{va} phase center position 
is computed by applying~\eqref{eq:posAnchor-SB} twice, i.e., $\posVA{j}{s}{s'} = \hva(\hva(\posAnchor{j}, \pmva{s}), \pmva{s'})$, which is exemplified in Fig.\,\ref{fig:SMC_incidence_angles}. 
The orientation transformation of the double bounce path $\MVApair{s}{s'}$ is taken into account by applying Householder matrices $\house{s}$ and $\house{s'}$ from both surfaces using~\eqref{eq:house} in \textit{inverse} order.
That is, with $\range{n}{j}{s}{s'}\!=\!\pos{n}\!-\!\posVA{j}{s}{s'}$ pointing from the double-bounce \ac{va} to the \ac{pm} in \textit{global} Cartesian coordinates, the vector pointing from \ac{pa} $j$ to the double-bounce \ac{vm} becomes $\ranget{n}{j}{s}{s'}\!=\house{s}\house{s'} \range{n}{j}{s}{s'}$. 
Both the vectors $\rangep{n}{j}{s}{s'}\!=\!\rotM^{-1} \big( \ranget{n}{j}{s}{s'} \big)$ and $\rangepRX{n}{j}{s}{s'}\!=\! \rotMn^{-1} \big( \!-\!\range{n}{j}{s}{s'} \big)$ in local Cartesian \ac{pa} and \ac{pm} coordinates, respectively, as well as the local channel parameters
in distance, \ac{aoa}, and \ac{aod}, compute analogously to the single-bounce case:
\begin{align}
    \dist{s}{s'}(\pos{n},\pmva{s},\pmva{s'}) &\!=\!\lVert \range{n}{j}{s}{s'} \rVert  \, , \label{eq:DB-dist} \\
	\aoa{s}{s'}(\pos{n},\pmva{s},\pmva{s'}) &\!=\! \arctantwo \big([\rangepRX{n}{j}{s}{s'}]_2,[\rangepRX{n}{j}{s}{s'}]_1\big) \, , \label{eq:DB-aoa}\\
	\aod{s}{s'}(\pos{n},\pmva{s},\pmva{s'}) &\!=\! \arctantwo \big([\rangep{n}{j}{s}{s'}]_2,[\rangep{n}{j}{s}{s'}]_1\big) \,. \label{eq:DB-aod}
\end{align}

%% file: figures/SMC_incidence_angles.tex
%
%
\definecolor{mycolor1}{rgb}{0.72941,0.04706,0.18431}%
\definecolor{mycolor2}{rgb}{0.88235,0.63922,0.00000}%
\definecolor{mycolor3}{rgb}{0.00000,0.38431,0.60784}%
\definecolor{mycolor4}{rgb}{0.52549,0.12157,0.25490}%
\definecolor{mycolor1x}{rgb}{0.00000,0.15686,0.33333}%
\definecolor{mycolor2x}{rgb}{0.00000,0.38431,0.60784}%
\definecolor{mycolor3x}{rgb}{0.88235,0.63922,0.00000}%
\definecolor{mycolor4x}{rgb}{0.72941,0.04706,0.18431}%
\definecolor{mycolor5x}{rgb}{0.52549,0.12157,0.25490}%

\pgfplotsset{every axis/.append style={
  label style={font=\footnotesize},
  legend style={font=\footnotesize},
  tick label style={font=\footnotesize},
}}

\pgfplotsset{every axis/.append style={
  label style={font=\footnotesize},
  legend style={font=\footnotesize},
  tick label style={font=\footnotesize},
  xticklabel={
    \ifdim \tick pt < 0pt
      \pgfmathparse{abs(\tick)}%
      \llap{$-{}$}\pgfmathprintnumber{\pgfmathresult}
   \else
      \pgfmathprintnumber{\tick}
   \fi
}}}

\begin{tikzpicture}

\begin{axis}[%
width=\plotWidth,
height=0.55\plotWidth,
at={(0\plotWidth,0\plotWidth)},
scale only axis,
xmin=-4.2,
xmax=5.35,
line cap = round,               
line join = round,               
ymin=-0.25,
ymax=5,
grid = none,        
ticks=none,         
axis lines=none,    
]

\input{./figures/SMC_arrays_only}



\addplot [color=black, only marks, mark=*, mark options={solid, mycolor3},  mark size=1pt, forget plot]
  table[row sep=crcr]{%
0	0\\
};
\node[centered, align=center, inner sep=0]
at (axis cs:0,-0.15) {$\scriptstyle\bm{0}$};
\addplot [color=black, forget plot, line width=1pt]
  table[row sep=crcr]{%
-1	-50\\
-1	50\\
};
\addplot [color=black, forget plot, line width=1pt]
  table[row sep=crcr]{%
-8.07898704092593	49.3670740436207\\
1.51344167848865	0.295847125364877\\
11.1058703979032	-48.7753797928909\\
};
\node[centered, align=center, inner sep=0]
at (axis cs:-0.85,0) {$s$};
\node[centered, align=center, inner sep=0]
at (axis cs:1.38,0) {$s'$};

\node[centered, align=center, inner sep=0]
at (axis cs:-0.6,4.4) {\footnotesize agent};
\node[left, align=right, inner sep=0]
at (axis cs:-0.6,4.0) {$\scriptstyle \pos{n}$};

\node[centered, align=center, inner sep=0,yshift=1mm]
at (axis cs:-3.3,4.736) {\footnotesize double-bounce VM};
\node[centered, align=center, inner sep=0]
at (axis cs:0.501,0.25) {\footnotesize PA};
\node[centered, align=center, inner sep=0]
at (axis cs:4.75,2.45) {\footnotesize double-bounce\\[-3pt]\footnotesize VA};

\node[left, align=right, inner sep=0]
at (axis cs:-2.7,0.5) {\footnotesize single-bounce\\[-3pt]\footnotesize VA};
\node[right, align=left, inner sep=0]
at (axis cs:2.2,4.5) {\footnotesize single-bounce\\[-3pt]\footnotesize VM};

\addplot [color=mycolor3, line width=1.2pt, only marks, mark=*,  mark size=1pt,  mark options={solid, mycolor3}, forget plot]
  table[row sep=crcr]{%
-2	0\\
};
\node[centered, align=center, inner sep=0]
at (axis cs:-1.999,0.21) {$\scriptstyle \pmva{s}$};
\addplot [color=mycolor3, line width=1.2pt, only marks,  mark size=1pt, mark=*, mark options={solid, mycolor3}, forget plot]
  table[row sep=crcr]{%
3.0268833569773	0.591694250729754\\
};
\node[centered, align=center, inner sep=0]
at (axis cs:3.028,0.802) {$\scriptstyle \pmva{s'}$};
\addplot[-stealth, color=mycolor4, dashed, line width=0.7pt, draw opacity = 0.4,%
{-Stealth[inset=0pt, scale=1.05, angle'=25]}
]
  table[row sep=crcr]{%
5.15456908094817	1.99631286801011\\
5.15456908094817	1.99631286801011\\
-0.5	4\\
};
\node[right, align=left, inner sep=0,color=mycolor4]
at (axis cs:-0.5,3.7) {$\scriptstyle \range{n}{j}{s}{s'}$};
\addplot[-stealth, color=mycolor4, dashed, line width=0.7pt, draw opacity = 0.4,%
{-Stealth[inset=0pt, scale=1.05, angle'=25]}
]
  table[row sep=crcr]{%
0.5	0.5\\
0.5	0.5\\
-3.98379829304221	4.48553214532623\\
};
\node[right, align=left, inner sep=0,color=mycolor4]
at (axis cs:-3.683,4.285) {$\scriptstyle \ranget{n}{j}{s}{s'}$};
\addplot [color=mycolor4, dashed, line width=0.7pt, forget plot, draw opacity = 0.4]
  table[row sep=crcr]{%
-2.5	0.5\\
1.98379829304221	4.48553214532623\\
};
\addplot [color=mycolor3, line width=0.7pt, forget plot, draw opacity = 0.4]
  table[row sep=crcr]{%
0.5	0.5\\
-1	1.8333111409731\\
0.885312695136672	3.5091167547356\\
-0.5	4\\
};
\addplot[-stealth,color=black, line width=0.7pt,%
{-Stealth[inset=0pt, scale=1.05, angle'=25]}
]
  table[row sep=crcr]{%
-1	1.8333111409731\\
0	1.8333111409731\\
};
\node[right, align=left, inner sep=0]
at (axis cs:0,1.833) {$\scriptstyle \nwc{s}$};
\addplot[-stealth,color=black, line width=0.7pt,%
{-Stealth[inset=0pt, scale=1.05, angle'=25]}
]
  table[row sep=crcr]{%
0.885312695136672	3.5091167547356\\
-0.0961118432284438	3.31726818034731\\
};
\node[left, align=right, inner sep=0]
at (axis cs:-0.05,3.28) {$\scriptstyle \nwc{s'}$};
\addplot [color=black, line width=0.7pt, forget plot]
  table[row sep=crcr]{%
-0.25	1.8333111409731\\
-0.250130173029195	1.8472840647464\\
-0.250520646930065	1.86125213812554\\
-0.251171286158158	1.87521051240004\\
-0.252081864858327	1.88915434222627\\
-0.253252066943139	1.90307878730939\\
-0.254681486202593	1.91697901408354\\
-0.25636962644513	1.93085019738972\\
-0.258315901669871	1.94468752215073\\
-0.26051963627004	1.95848618504262\\
-0.26298006526748	1.97224139616207\\
-0.265696334578203	1.98594838068908\\
-0.268667501308864	1.99960238054446\\
-0.271892534084067	2.01319865604149\\
-0.275370313404383	2.02673248753117\\
-0.279099632034963	2.04019917704062\\
-0.283079195424597	2.05359404990377\\
-0.287307622155096	2.06691245638416\\
-0.291783444420816	2.08014977328895\\
-0.296505108538179	2.09330140557376\\
-0.301470975484993	2.10636278793776\\
-0.306679321469412	2.1193293864084\\
-0.312128338528304	2.13219669991526\\
-0.317816135154853	2.14496026185256\\
-0.323740736955147	2.15761564162955\\
-0.329900087333552	2.1701584462086\\
-0.33629204820661	2.18258432163005\\
-0.342914400745229	2.19488895452365\\
-0.349764846144903	2.20706807360584\\
-0.356841006423686	2.21911745116243\\
-0.364140425247664	2.23103290451613\\
-0.371660568783609	2.24281029747853\\
-0.379398826578551	2.25444554178586\\
-0.387352512465936	2.26593459851811\\
-0.395518865498071	2.27727347950111\\
-0.403895050904523	2.2884582486909\\
-0.412478161076153	2.29948502354005\\
-0.421265216574425	2.31034997634539\\
-0.430253167165654	2.32104933557675\\
-0.439438892879827	2.33157938718613\\
};
\addplot [color=black, line width=0.7pt, forget plot]
  table[row sep=crcr]{%
-0.25	1.8333111409731\\
-0.250130173029195	1.8193382171998\\
-0.250520646930065	1.80537014382066\\
-0.251171286158158	1.79141176954616\\
-0.252081864858327	1.77746793971993\\
-0.253252066943139	1.76354349463681\\
-0.254681486202593	1.74964326786266\\
-0.25636962644513	1.73577208455648\\
-0.258315901669871	1.72193475979547\\
-0.26051963627004	1.70813609690358\\
-0.26298006526748	1.69438088578413\\
-0.265696334578203	1.6806739012571\\
-0.268667501308864	1.66701990140174\\
-0.271892534084067	1.65342362590471\\
-0.275370313404383	1.63988979441503\\
-0.279099632034963	1.62642310490558\\
-0.283079195424597	1.61302823204243\\
-0.287307622155096	1.59970982556204\\
-0.291783444420816	1.58647250865725\\
-0.296505108538179	1.57332087637244\\
-0.301470975484993	1.56025949400844\\
-0.306679321469412	1.5472928955378\\
-0.312128338528304	1.53442558203094\\
-0.317816135154853	1.52166202009364\\
-0.323740736955147	1.50900664031665\\
-0.329900087333552	1.4964638357376\\
-0.33629204820661	1.48403796031615\\
-0.342914400745229	1.47173332742255\\
-0.349764846144903	1.45955420834036\\
-0.356841006423686	1.44750483078377\\
-0.364140425247664	1.43558937743007\\
-0.371660568783609	1.42381198446767\\
-0.379398826578551	1.41217674016034\\
-0.387352512465936	1.40068768342809\\
-0.395518865498071	1.38934880244509\\
-0.403895050904523	1.3781640332553\\
-0.412478161076153	1.36713725840615\\
-0.421265216574425	1.35627230560081\\
-0.430253167165654	1.34557294636945\\
-0.439438892879827	1.33504289476007\\
};

\addplot [color=black, line width=0.7pt, forget plot]
  table[row sep=crcr]{%
0.149244291362835	3.36523032394438\\
0.151281741297333	3.3551733924048\\
0.153456592788646	3.34514527716722\\
0.15576843873135	3.33514785537135\\
0.158216846376352	3.32518299841145\\
0.160801357411884	3.31525257158606\\
0.163521488049304	3.30535843374879\\
0.166376729113648	3.29550243696045\\
0.169366546138943	3.28568642614227\\
0.172490379468254	3.27591223873062\\
0.175747644358445	3.26618170433304\\
0.179137731089633	3.25649664438576\\
0.182660005079322	3.24685887181276\\
0.186313807001188	3.23727019068639\\
0.190098452908501	3.22773239588971\\
0.194013234362145	3.21824727278046\\
0.198057418563231	3.20881659685691\\
0.202230248490271	3.19944213342548\\
0.206530943040882	3.19012563727028\\
0.210958697177995	3.18086885232471\\
0.215512682080553	3.17167351134492\\
0.220192045298652	3.16254133558552\\
0.224995910913114	3.15347403447739\\
0.229923379699442	3.14447330530764\\
0.234973529296148	3.13554083290196\\
0.240145414377409	3.1266782893092\\
0.245438066830016	3.1178873334884\\
0.250850495934595	3.10916961099825\\
0.256381688551059	3.10052675368906\\
0.262030609308252	3.09196037939732\\
0.267796200797761	3.0834720916428\\
0.273677383771848	3.07506347932847\\
0.279673057345472	3.06673611644302\\
0.285782099202362	3.05849156176626\\
0.292003365805099	3.05033135857731\\
0.298335692609176	3.04225703436572\\
0.30477789428098	3.03427010054559\\
0.311328764919678	3.02637205217258\\
0.317987078282945	3.01856436766411\\
0.324751588016499	3.01084850852257\\
};
\addplot [color=black, line width=0.7pt, forget plot]
  table[row sep=crcr]{%
0.149244291362835	3.36523032394438\\
0.147344624370692	3.3753141892522\\
0.145583095915176	3.38542310075284\\
0.143960035732725	3.39555516618253\\
0.142475747640221	3.40570848894337\\
0.14113050947812	3.41588116845836\\
0.139924573058442	3.42607130052713\\
0.138858164117638	3.43627697768244\\
0.13793148227433	3.44649628954717\\
0.137144700991951	3.45672732319197\\
0.136497967546268	3.46696816349331\\
0.13599140299782	3.47721689349198\\
0.135625102169254	3.48747159475191\\
0.135399133627574	3.49773034771928\\
0.13531353967131	3.50799123208185\\
0.135368336322596	3.5182523271284\\
0.135563513324173	3.52851171210828\\
0.13589903414131	3.53876746659094\\
0.136374835968641	3.54901767082542\\
0.136990829741923	3.55926040609968\\
0.137746900154705	3.56949375509981\\
0.138642905679915	3.57971580226885\\
0.139678678596351	3.58992463416544\\
0.140854025020076	3.60011833982196\\
0.142168724940711	3.6102950111022\\
0.143622532262618	3.6204527430586\\
0.145215174850966	3.63058963428881\\
0.146946354582672	3.6407037872916\\
0.148815747402206	3.65079330882206\\
0.150823003382248	3.66085631024599\\
0.152967746789194	3.67089090789342\\
0.155249576153484	3.68089522341121\\
0.157668064344757	3.69086738411469\\
0.1602227586518	3.70080552333816\\
0.162913180867295	3.7107077807843\\
0.165738827377329	3.72057230287246\\
0.168699169255664	3.73039724308554\\
0.171793652362751	3.74018076231572\\
0.175021697449454	3.74992102920866\\
0.178382700265476	3.75961622050631\\
};

\addplot [color=black, line width=1.2pt, only marks, mark=*,  mark size=0.5pt,  mark options={solid, black}, forget plot]
  table[row sep=crcr]{%
5.1546	1.9963\\
};
\addplot[color=black, line width=0.35pt, draw opacity = 0.9]
  table[row sep=crcr]{%
5.1546	1.9963\\
4.6546	1.4963\\
4.1546	1.4963\\
};
\node[left, align=right, inner sep=0]
at (axis cs:4.1546,1.4963) {$\scriptstyle \posVA{j}{s}{s'}$};

\addplot [color=black, line width=1.2pt, only marks, mark=*,  mark size=0.5pt,  mark options={solid, black}, forget plot]
  table[row sep=crcr]{%
0.5	0.5\\
};
\addplot[color=black, line width=0.35pt, draw opacity = 0.9]
  table[row sep=crcr]{%
0.5	0.5\\
0.7	0.3\\
0.9	0.3\\
};
\node[right, align=left, inner sep=0]
at (axis cs:0.9,0.3) {$\scriptstyle \posAnchor{j}$};

\addplot[area legend, line width=1.2pt, draw=none, fill=mycolor3, fill opacity=0.035, forget plot]
table[row sep=crcr] {%
x	y
0.5	0.5\\
-1	1.83337\\
0.8853	3.5091\\
0.5	0.5\\
}--cycle;

\addplot[area legend, line width=1.2pt, draw=none, fill=mycolor3, fill opacity=0.035, forget plot]
table[row sep=crcr] {%
x	y
-0.5	4\\
-1	1.83337\\
0.8853	3.5091\\
-0.5	4\\
}--cycle;

\addplot[-stealth, line width=0.7pt, draw opacity = 0.9,%
{-Stealth[inset=0pt, scale=1.05, angle'=25]}
]
  table[row sep=crcr]{%
0.48	0.48\\
-0.117931847594851	1.0114861292939\\
};
\node[left, align=right, inner sep=0]
at (axis cs:-0.098,1.031) {$\scriptstyle \ert{n}{j}{s}{s'}$};
\addplot[-stealth, line width=0.7pt, draw opacity = 0.9,%
{-Stealth[inset=0pt, scale=1.05, angle'=25]}
]
  table[row sep=crcr]{%
5.13456908094817	1.97631286801011\\
4.38051041975223	2.24351229816553\\
};
\node[below, align=right, inner sep=0]
at (axis cs:4.301,2.2) {$\scriptstyle \er{n}{j}{s}{s'}$};

\end{axis}
\end{tikzpicture}%

%% file: InputFiles/SystemModel.tex
At each time $ n $, the state of the mobile agent is given by $ \RV{x}_{n} := [\RV{p}_{n}^{\mathrm{T}} \iist \RV{v}_{n}^{\mathrm{T}} \iist \rv{{\Delta\varphi}}_{n}]^{\mathrm{T}}$ consisting of the position $ \RV{p}_{n} $, the velocity $ \RV{v}_{n} =[\rv{v}_{\mathrm{x},n} \iist \rv{v}_{\mathrm{y},n}]^{\mathrm{T}} $ and the orientation $ \rv{{\Delta\varphi}}_{n}$. 
All agent states up to time $ n $ are denoted as $ \RV{x}_{1:n} := [\RV{x}_{1}^{\mathrm{T}} \ist\cdots\ist \RV{x}_{n}^{\mathrm{T}} ]^{\mathrm{T}} $. 
Impinging on the agent, each component $\MVApair{s}{s'}$ 
 has parameters $\RV{{\beta}}_{ss',n}^{(j)} \!:=\! [ \rv{{u}}_{ss',n}^{(j)} \iist \rv{r}_{ss',n}^{(j)}]^{\mathrm{T}}$ consisting of the normalized amplitude 
 (the square root of the component SNR~\cite{XuhongTWC2022})
 $\rv{u}_{ss',n}^{(j)}$ and the binary existence variable $\rv{r}_{ss',n}^{(j)} \in \{0,1\} $ that is to be understood as visibility (cf.\,\cite{wilding2023propagation}), meaning that component $\MVApair{s}{s'}$ is visible if and only if ${r}_{ss',n}^{(j)} = 1$ .

\subsection{Measurement Likelihood Functions}

The agent state, \ac{surface} positions, and component parameters relate to the distance measurements ${\rv{z}_\mathrm{d}}_{m,n}^{(j)}$, the \ac{aoa} measurements ${\rv{z}_\mathrm{\varphi}}_{m,n}^{(j)}$, the \ac{aod} measurements $ {\rv{z}_\mathrm{\phi}}_{m,n}^{(j)} $, and the normalized amplitude measurements ${\rv{z}_\mathrm{u}}_{m,n}^{(j)} $ via the following \acp{lhf}, which are assumed to be conditionally independent of each other. 
The individual \acp{lhf} of the distance, \ac{aoa}, and \ac{aod} measurements are modeled by Gaussian \acp{pdf}
\begin{align}
	f_{ss'}^{(j)}({z_\mathrm{d}}_{m,n}^{(j)}) := f_{\mathrm{N}}\big({z_\mathrm{d}}_{m,n}^{(j)}; d_{ss',n}^{(j)}, \sigma_{\mathrm{d}}(u_{ss',n}^{(j)})\big), 
	\label{eq:LHF_dist} \\[-7mm]\nn
\end{align}
\begin{align}
	f_{ss'}^{(j)}({z_\mathrm{\varphi}}_{m,n}^{(j)}) := f_{\mathrm{N}}\big({z_\mathrm{\varphi}}_{m,n}^{(j)} ; \varphi_{ss',n}^{(j)}, \sigma_{\mathrm{\varphi}}(u_{ss',n}^{(j)})\big), 
	\label{eq:LHF_AoA}\\[-7mm]\nn
\end{align}
\begin{align}
	f_{ss'}^{(j)}({z_\mathrm{\phi}}_{m,n}^{(j)}) := f_{\mathrm{N}}\big({z_\mathrm{\phi}}_{m,n}^{(j)} ; \phi_{ss',n}^{(j)}, \sigma_{\mathrm{\phi}}(u_{ss',n}^{(j)})\big), 
	\label{eq:LHF_AoD}  \\[-4mm]\nn
\end{align}
and apply for \ac{los} paths, single-bounce paths, and double-bounce paths, depending on whether $(s\rmv,\rmv s') \!=\! (0,0)$, $(s\rmv,\rmv s') \!\in\! \setSB$, or $(s\rmv,\rmv s') \!\in\! \setDB$. 
The means $d_{ss',n}^{(j)}$, $\phi_{ss',n}^{(j)}$, and $\varphi_{ss',n}^{(j)}$ are calculated according to Sec.\,\ref{sec:GeometricalRelations}. 
The variances of the Gaussian \acp{pdf} $ {\sigma_\mathrm{d}}^2(u_{ss',n}^{(j)}) $, $ \sigma_{\mathrm{\phi}}^2(u_{ss',n}^{(j)}) $, and $ \sigma_{\mathrm{\varphi}}^2(u_{ss',n}^{(j)}) $ depend on the normalized amplitude $u_{ss',n}^{(j)}$ and are determined based on the Fisher information through 
\begin{align} 
{\sigma_\mathrm{d}}^2(u_{ss',n}^{(j)}) &= \lightspeed^2/\big(8\pi^2 \beta_{\mathrm{bw}}^2 (u_{ss',n}^{(j)})^2\big) \,, \label{eq:varianceDist} \\
 \sigma_{\varphi}^2(u_{ss',n}^{(j)}) &= \lightspeed^2/\big(8\pi^2f_{\mathrm{c}}^2 (u_{ss',n}^{(j)})^2 D^2(\varphi_{ss',n}^{(j)})\big) \,, \label{eq:varianceAoA} \\
 \sigma_{\phi}^2(u_{ss',n}^{(j)}) &= \lightspeed^2/\big(8\pi^2f_{\mathrm{c}}^2 (u_{ss',n}^{(j)})^2 D^2(\phi_{ss',n}^{(j)})\big) \,, \label{eq:varianceAoD}
\end{align} 
with $\lightspeed$ and $f_{\mathrm{c}}$ denoting the speed of light and the carrier frequency, respectively, $ \beta_{\mathrm{bw}}^2$ denoting the mean square bandwidth of the transmit signal pulse and $ D^2(\cdot) $ being the squared array aperture \cite{Thomas_Asilomar2018, LeitingerICC2019}.
The \ac{lhf} of the normalized amplitude is denoted $f_{ss'}^{(j)}(z_{\mathrm{u}_{m,n}}^{(j)})$. 
Using \eqref{eq:LHF_dist} through \eqref{eq:varianceAoD}, the \acp{lhf} for 
\ac{los}, single-, and double-bounce paths, respectively, are factorized as follows
\begin{align}
	& f\big(\V{z}_{m,n}^{(j)}|\V{x}_{n}, u_{00,n}^{(j)}\big) \label{eq:LHF_PA} \\ 
	& \hspace*{5mm} =  f_{00}^{(j)}({z_\mathrm{d}}_{m,n}^{(j)}) f_{00}^{(j)}({z_\mathrm{\phi}}_{m,n}^{(j)}) f_{00}^{(j)}({z_\mathrm{\varphi}}_{m,n}^{(j)}) f_{00}^{(j)}({z_\mathrm{u}}_{m,n}^{(j)}), 	 \nn \\
	& f\big(\V{z}_{m,n}^{(j)}|\V{x}_{n}, \V{p}_{s,\mathrm{\surface}}^{(j)}, u_{ss,n}^{(j)}\big) \label{eq:LHF_Spath} \\ 
	& \hspace*{5mm} =  f_{ss}^{(j)}({z_\mathrm{d}}_{m,n}^{(j)}) f_{ss}^{(j)}({z_\mathrm{\phi}}_{m,n}^{(j)}) f_{ss}^{(j)}({z_\mathrm{\varphi}}_{m,n}^{(j)}) f_{ss}^{(j)}({z_\mathrm{u}}_{m,n}^{(j)}), 
	\nn \\
	& f\big(\V{z}_{m,n}^{(j)}|\V{x}_{n}, \V{p}_{s,\mathrm{\surface}}^{(j)}, \V{p}_{s',\mathrm{\surface}}^{(j)}, u_{ss',n}^{(j)}\big) \label{eq:LHF_Dpath}  \\ 
	& \hspace*{5mm} =  f_{ss'}^{(j)}({z_\mathrm{d}}_{m,n}^{(j)}) f_{ss'}^{(j)}({z_\mathrm{\phi}}_{m,n}^{(j)}) f_{ss'}^{(j)}({z_\mathrm{\varphi}}_{m,n}^{(j)}) f_{ss'}^{(j)}({z_\mathrm{u}}_{m,n}^{(j)}). \nn
\end{align}
Before being observed, the measurements $ \RV{z}_{m,n}^{(j)} \rmv\!:=\!\rmv [ {\rv{z}_\mathrm{d}}_{m,n}^{(j)} \iist {\rv{z}_\mathrm{\phi}}_{m,n}^{(j)} \iist {\rv{z}_\mathrm{\varphi}}_{m,n}^{(j)} \iist {\rv{z}_\mathrm{u}}_{m,n}^{(j)} ]^{\mathrm{T}} \rmv\in\rmv \mathbb{R}^{4\rmv\times\rmv1} $ with $m \in \{1,\dots,\rv{M}_{n}^{(j)}\}$ and the measurement number $\rv{M}_{n}^{(j)}$ at each \ac{pa} are considered as random. The joint measurement vectors for all \acp{pa} and all times up to $n$ are given by  $ \RV{z}_{n}^{(j)} \rmv\!:=\!\rmv [\RV{z}_{1,n}^{{(j)}\mathrm{T}} \ist \cdots \ist \RV{z}_{\rv{M}_{n}^{(j)}\!\!,n}^{^{(j)}\mathrm{T}}]^{\mathrm{T}} \rmv\in\rmv \mathbb{R}^{4\rv{M}_{n}^{(j)}\rmv\times\rmv1}$, $ \RV{z}_{n} \rmv\!:=\!\rmv [\RV{z}_{n}^{{(1)}\mathrm{T}} \ist \cdots \ist \RV{z}_{n}^{{(J)}\mathrm{T}}]^{\mathrm{T}} $ and $ \RV{z}_{1:n} \rmv\!:=\!\rmv [\RV{z}_{1}^{\mathrm{T}} \ist \cdots \ist \RV{z}_{n}^{\mathrm{T}}]^{\mathrm{T}} $. 
In practice, the measurements are obtained by applying a snapshot-based channel estimation and detection algorithm to the observed discrete \ac{rf} signals.

%% file: InputFiles/PCRLB.tex
Let $\RVetaglobal{n}\!:=\![ \RVstate{n}^\trp, \RVpMVAposStacked^\trp]^\trp \! \in \!\realset{\dimGlobal}{1}$ denote the joint state vector of dimension $\dimGlobal\!=\!5+2S$ comprising the agent state $\RVstate{n}\!\in \!\realset{5}{1}$ and the stacked \ac{surface} positions $\RVpMVAposStacked\!:=\![\RVpmva{1}^\trp \dots \RVpmva{S}^\trp]^\trp \!\in\! \realset{2S}{1}$ which define the environment map.
Assuming a nearly constant velocity model~\cite[Sec.\,6.3.2]{BarShalom_Tracking2004} the evolution of the joint state is defined through the state-transition model
\begin{align}\label{eq:state-transition-model}
	\RVetaglobal{n} = \transitionmatrix \RVetaglobal{n-1} + \RVprocessNoise{n} 
\end{align}
where $f(\processNoise{n}):=\pdfGaussian(\processNoise{n};\bm{0},\processNoiseCov)$.
The state-transition matrix is 
\begin{align}
    \left[\transitionmatrix\right]_{\scriptscriptstyle k,\ell} = 
    \begin{cases}
        1, & k=\ell \\
        \mathrm{T}, & \big((k,\ell) = (1,3)\big) \lor \big((k,\ell) = (2,4)\big) \\
        0, & \text{else}
    \end{cases},
\end{align}
with $\mathrm{T}$ denoting the time interval between two subsequent observations $\observationMatrix{n-1}$ and $\observationMatrix{n}$.
The process noise covariance matrix $\processNoiseCov \in \realset{\dimGlobal}{\dimGlobal}$ is 
\begin{align}
    \left[\processNoiseCov \right]_{\scriptscriptstyle k,\ell} = 
    \begin{cases}
         \left[ \noiseVarianceV \gainmatrix  \gainmatrix^\trp \right]_{\scriptscriptstyle k,\ell}, & \big(k \leq 4\big) \land \big( \ell \leq 4\big)\\
        \noiseVarianceO, & \big(k= 5\big) \land \big( \ell = 5\big)\\
        \noiseVarianceP, & \big(k = \ell \big) \land \big(k \geq 6\big) \\ 
        0, & \text{else}
    \end{cases},
\end{align}
where $\noiseVarianceV$ denotes the process noise variance of the kinematic agent state excluding orientation,
$\noiseVarianceO$ denotes the process noise variance of the agent orientation, 
and $\noiseVarianceP$ denotes the process noise variance of \ac{surface} positions (accounting for wall non-idealities such as curvature).
The gain matrix is
\begin{align}
	\gainmatrix = 
		\begin{bmatrixs}
			\frac{\mathrm{T}^2}{2} & 0 \\
			0 & \frac{\mathrm{T}^2}{2} \\
			\mathrm{T} & 0 \\
			0 & \mathrm{T} \\
		 \end{bmatrixs}\,.
\end{align}

We are ultimately interested in obtaining the global~\gls{pcrlb}
\begin{align}\label{eq:posteriorCRLB}
	\PCRLB = 
	\Big(
		\underbrace{\FIMglobal{n} + \FIMstep{n}{n\!-\!1}}_{=:\FIMstep{n}{n}}
	\Big)^{-1}
\end{align}
that is a lower bound on the \ac{mse} matrix~\cite[eq.\,(29)]{VanTrees2007PCRLB} 
of any 
estimator%
\footnote{For any square matrix $\M{X}$, the notation $\M{X} \succeq \M{0}$ is to be interpreted as $\M{X}$ being positive semidefinite~\cite{Kay_EstimationTheory}.} 
$\mathbb{E}_{\RVetaglobal{0:n} , \RVobservationMatrix{1:n}}\!\!\big((\RVetaglobalHat{n}\!-\!\RVetaglobal{n} ) (\RVetaglobalHat{n}\!-\!\RVetaglobal{n} )^\trp\big) \succeq \PCRLB$.
The \ac{pcrlb} matrix $\PCRLB$ is the inverse of the posterior~\gls{fim} $\FIMstep{n}{n}$ that is computed through the information fusion of the \gls{fim} $\FIMglobal{n}$ about the global parameters of interest $\RVetaglobal{n}$ obtained from a snapshot of observations $\observationMatrix{n}$ at the current time step $n$ with the predicted \gls{fim}
\footnote{%
Deriving the~\gls{fim} for the joint PDF $f(\etaglobal{0:n+1},\observationMatrix{1:n})$ and applying the Schur complement, Tichavsk\'y et al.
~\cite[eq.\,(21)]{Tichavsky98PCRLB} obtain $\FIMstep{n+1}{n}$. 
Under the linear Gaussian state-transition model~\eqref{eq:state-transition-model} Hernandez et al.~\cite{Hernandez02PCRLB} apply the Woodbury identity to arrive at the result in (25).
}
~\cite[eq.\,(16)]{Hernandez02PCRLB}
\begin{align}\label{eq:priorFIM}
	\FIMstep{n}{n\!-\!1} = 
	\left(
		\transitionmatrix \, \FIMstep{n\!-\!1}{n\!-\!1}^{-1} \, \transitionmatrix^\trp + \processNoiseCov
	\right)^{-1}
\end{align}
which 
is obtained by propagating 
the inverse of the old posterior \gls{fim} $\FIMstep{n\!-\!1}{n\!-\!1}$ over the state-transition model in~\eqref{eq:state-transition-model}.

\subsection{Global Per-Snapshot FIM}\label{sec:global-FIM} 

We derive the global per-snapshot \gls{fim} $\FIMglobal{n}$ under the assumption \nextassumption of correct detection (i.e., perfect inference of existences $\rv{r}_{ss',n}^{(j)}$) and data association 
(i.e., perfect association between measurements $m$ and \acp{mpc} $\MVApair{s}{s'}$)
of all components $\ncomponent\!\in\!\{1\dots\Ncomponents\}$ at time $n$. 
We also assume \nextassumption[ass:amplitudes] that the signal amplitudes $\rv{u}_{ss',n}^{(j)}$ contribute negligible information about the parameters of interest (i.e., the curvature of $\ln f_{ss'}^{(j)}(z_{\mathrm{u}_{m,n}}^{(j)})$ w.r.t. the joint state $\RVetaglobal{n}$ is low compared to the other \acp{lhf}) and we treat them as deterministic knowns in this \ac{fim} analysis. 
As a result, the variances $\varianceDist$, $\varianceAoA$, and $\varianceAoD$ in~\eqref{eq:varianceDist}-\eqref{eq:varianceAoD} become deterministic functions in ${u}_{ss',n}^{(j)} $, when conditioned on the channel parameters $\RVetach{n}{j}$ defined below. 
We further assume \nextassumption[ass:variances] that the variances $\varianceDist$, $\varianceAoA$, and $\varianceAoD$ contribute negligible information about these channel parameters (i.e., $\etach{n}{j}$) and we treat them as deterministic knowns. 
Under these assumptions, no nuisance parameters enter the per-anchor observation likelihood.
Assuming \nextassumption that each anchor $j$ contributes independent information on $\etaglobal{n}$, i.e., assuming conditionally independent observations $\observation{n}{j}\!\!$, the \gls{fim} for the global parameter vector is%
\footnote{%
The elements 
$[\FIMglobal{n}]_{k,\ell}\!=\!\mathbb{E}_{\RVetaglobal{0:n} , \RVobservationMatrix{1:n}}\!\!\big(\frac{-\partial^2}{\partial \theta_k \partial \theta_\ell} \ln f(\observationMatrix{n}| \etaglobal{n}  )\big)$ are to be computed as the expectation under the joint PDF $f(\etaglobal{0:n} , \observationMatrix{1:n})$ which can be shown to correspond to the expectation under the prior PDF $f(\etaglobal{n})$ of the ``classic'' FIM (computed as the expectation under the likelihood $f(\observationMatrix{n}|\etaglobal{n})$). 
Because the resulting expression $\FIMglobal{n} = \mathbb{E}_{\RVetaglobal{n} } \!\!\big(\sum_j \jacobgn{j}(\etaglobal{n}) \FIMch{n}{j} \jacobgn{j}^\trp(\etaglobal{n})\big)$ is analytically intractable, we concentrate 
$f(\etaglobal{n}) \approx \delta(\etaglobal{n}\!-\!{\etaglobal{n}}^{\!\star})$ at the ground truth state ${\etaglobal{n}}^{\!\star}$. 
}\cite{Thomas_Asilomar2018}
\begin{align}\label{eq:FIMglobal}
    \FIMglobal{n} \approx \sum\limits_{j=1}^{J} \jacobgn{j} \FIMch{n}{j} \jacobgn{j}^\trp \qquad \in  \realset{\dimGlobal}{\dimGlobal}
\end{align}
which is the sum of the local channel \glspl{fim} $\FIMch{n}{j} \in \realset{\dimLocal}{\dimLocal}$ contributed by all $J$ anchors, propagated via the Jacobian matrices $\jacobgn{j}:= \nicefrac{\partial {\etach{n}{j}}^{\!\!\!\trp}}{\partial \etaglobal{n}} \in  \realset{\dimGlobal}{\dimLocal}$ from local channel parameter level to global parameter level.

	\tikzset{
  		every picture/.append style={
    		line cap=round,
    		line join=round}
	}
	\newcommand{\MCruns}{ {100} }				
\newcommand{\OSPAcutoff}{ {\SI{5}{\metre}} }				
\newcommand{\LWbound}{1pt}
\newcommand{\LWestimates}{0.5pt}
\definecolor{mycolor1}{rgb}{0.00000,0.38431,0.60784}%
\definecolor{mycolor2}{rgb}{0.72941,0.04706,0.18431}%
\definecolor{mycolor3}{rgb}{0.00000,0.51765,0.23922}%
\definecolor{mycolor4}{rgb}{0.59608,0.11373,0.59216}%
\definecolor{mycolor5}{rgb}{0.88235,0.63922,0.00000}%
\colorlet{mycolor1}{IEEEblue}
\colorlet{mycolor2}{IEEEred}
\colorlet{mycolor3}{IEEEgreen}
\colorlet{mycolor4}{IEEEviolett}
\colorlet{mycolor5}{IEEEorange}

\subsection{Local Per-Anchor Channel FIM}\label{sec:local-FIM} 

The local channel parameter vectors 
\begin{align}
    \etach{n}{j} = 
        \left[ 
			{\distVec}^\trp\!\!,
			{\aoaVec}^\trp\!\!,
			{\aodVec}^\trp
        \right]^\trp \qquad \in \realset{\dimLocal}{1}
\end{align}
contain the stacked distances 
$\distVec:=[\dist{s}{s'}]_{\MVApair{s}{s'}\in\setSDBt} \in \realset{\Ncomponents}{1}$,
\acp{aoa} $\aoaVec:=[\aoa{s}{s'}]_{\MVApair{s}{s'}\in\setSDBt} \in \realset{\Ncomponents}{1}$,
and \acp{aod} $\aodVec:=[\aod{s}{s'}]_{\MVApair{s}{s'}\in\setSDBt} \in \realset{\Ncomponents}{1}$,
regardless of whether a measurement exists ${r}_{ss',n}^{(j)}=1$ or not ${r}_{ss',n}^{(j)}=0$.
The dimensionality of the channel parameter vector is $\dimLocal = 3 \Ncomponents\!$.

For brevity, we avoid the detection problem in this \ac{fim} analysis that would introduce random numbers $\rv{M}_{n}^{(j)}$ of measurements per \ac{pa} $j$ and use the constant maximum number $\Ncomponents$ of measurements
$ \RV{z}_{{s}{s'},n}^{(j)} \rmv\triangleq\rmv [ {\rv{z}_\mathrm{d}}_{{s}{s'},n}^{(j)} \iist {\rv{z}_\mathrm{\phi}}_{{s}{s'},n}^{(j)} \iist {\rv{z}_\mathrm{\varphi}}_{{s}{s'},n}^{(j)} \iist {\rv{z}_\mathrm{u}}_{{s}{s'},n}^{(j)} ]^{\mathrm{T}} \in \mathbb{R}^{4\rmv\times\rmv1} $ regardless of whether a measurement exists, 
and absorb the measurement existences ${r}_{ss',n}^{(j)}$ into~\eqref{eq:channel-fim-definition}.
Under the assumptions \nextassumption that the \acp{lhf} in \eqref{eq:LHF_PA}--\eqref{eq:LHF_Dpath} factorize and \nextassumption[ass:meas-uncorrelated] that the measurements $\RV{z}_{{s}{s'},n}^{(j)}$ and $\RV{z}_{\tilde{s}\tilde{s}',n}^{(j)}$ 
are uncorrelated for $\MVApair{s}{s'} \neq \MVApair{\tilde{s}}{\tilde{s}'}$, the channel \ac{fim} $\FIMch{n}{j}$ is diagonal.
Let 
\begin{subequations}
\begin{align}
  \iota_{d}\colon\;\setSDBt &\longrightarrow \{1 \dots \Ncomponents\},\\
  \iota_{\varphi}\colon\;\setSDBt &\longrightarrow \{\Ncomponents+1 \dots 2\Ncomponents\},\\	
  \iota_{\phi}\colon\;\setSDBt &\longrightarrow \{2\Ncomponents+1 \dots 3\Ncomponents\},
\end{align}
\end{subequations}
define the bijective mappings from the set of components $\setSDBt$ to the measurements such that $[\etach{n}{j}]_{\iota_{d}\MVApair{s}{s'}} = \dist{s}{s'}$, $[\etach{n}{j}]_{\iota_{\varphi}\MVApair{s}{s'}} = \aoa{s}{s'}$, and $[\etach{n}{j}]_{\iota_{\phi}\MVApair{s}{s'}} = \aod{s}{s'}$.
Under assumptions \assref{ass:amplitudes} through \assref{ass:meas-uncorrelated}, given the Gaussian observation \acp{lhf} in \eqref{eq:LHF_dist}--\eqref{eq:LHF_AoD}, and absorbing the existences ${r}_{ss',n}^{(j)}$, the per-anchor channel \ac{fim} becomes
\begin{align}\label{eq:channel-fim-definition}
    \left[\FIMch{n}{j}\right]_{\grave{\imath},\acute{\imath}} &= 
	\begin{cases}
			\frac{{r}_{ss',n}^{(j)}}{ \varianceDist(u_{ss',n}^{(j)})} & \grave{\imath}=\acute{\imath}=\iota_d(s,s')\\
			\frac{{r}_{ss',n}^{(j)}}{ \varianceAoA(u_{ss',n}^{(j)})} & \grave{\imath}=\acute{\imath}=\iota_\varphi(s,s') \\ 
			\frac{{r}_{ss',n}^{(j)}}{ \varianceAoD(u_{ss',n}^{(j)})} & \grave{\imath}=\acute{\imath}=\iota_\phi(s,s') \\
            0 & \text{otherwise.}
	\end{cases}    
\end{align}
This diagonal channel \ac{fim} model in~\eqref{eq:channel-fim-definition} neglects correlations in the channel parameters.
While we use this model to match our algorithm in~\cite{XuhongTWC2025}, the rest of our derivations also hold when inserting a non-diagonal channel \ac{fim} $\FIMch{n}{j}$ for noisy, correlated channel observations such as in~\cite{Deutschmann26TSP,Fascista25RadioStripes}.

\def\datapath{.}
\begin{figure}[!t]
	\centering
	\hspace{-5mm}\scalebox{0.92}{\input{\datapath/figures/results/synResultColorBar.tex}}\label{subfig:synResultColorBar}\\[-2.55mm]
	\hspace{1.45mm}\pgfplotslegendfromname{legendFig_synResultPAs}\\[0mm]
	\vspace{-4mm}
	\hspace{-11mm}\subfloat[]{\scalebox{0.92}{\input{\datapath/figures/results/synResultPA1fig3.tex}}\label{subfig:synResultPA1fig3}} 
	\captionsetup[subfigure]{oneside,margin={0.85cm,0cm}}
	\hspace{-5mm}\subfloat[]{\scalebox{0.92}{\input{\datapath/figures/results/synResultPA2fig3.tex}}\label{subfig:synResultPA3fig3}}\\[1mm] 
	\captionsetup[subfigure]{oneside,margin={0.85cm,0cm}} 
	\vspace{-2mm}
	\caption{Simulation results from~\cite{XuhongTWC2025} with ground truths and estimates of reflecting surfaces, propagation paths and agent positions are shown at time step $n\!=\!18$. 
	The line representations of estimated surfaces are computed using the MMSE estimates of the detected \acp{surface}. 
	Estimated propagation paths are augmented with colors matching the colorbar and representing MMSE estimates of the \acp{snr}.}	 
	\label{fig:synResultTwoPAs}
\end{figure}

Mapping from local channel parameters to global parameters, the 
Jacobian matrices are defined as 
\begin{align}\label{eq:jacobian-main}
    \jacobgn{j} =   
    \frac{\partial \etach{n}{j}\!^\trp }{\partial \etaglobal{n}} =
    \begin{bmatrixs}
        \jacobP{d} & \jacobP{\varphi}& \jacobP{\phi} 	\\
        \bm{0} 					& \bm{0} 					& \bm{0} 	 \\
        \bm{0} 					& \jacobO{\varphi} 					& \bm{0} 	 \\
        \jacobM{d}& \jacobM{\varphi}& \jacobM{\phi}	\\
\end{bmatrixs} \quad \in \realset{\dimGlobal}{\dimLocal}
\end{align}
with submatrices defined in Appendix~\ref{sec:app-global-FIM}.
Information about the agent velocity $\vel{n}$ in the posterior \gls{fim} $\FIMstep{n}{n}$ of~\eqref{eq:posteriorCRLB} is obtained only by propagating Fisher information about the position over the state-space model in~\eqref{eq:priorFIM}. 
%
%
%
From the posterior \gls{fim} $\FIMstep{n}{n}$, we define the \gls{peb} as
$%
    {\PEB}_{\!,n} \!:=\! \sqrt{\tr \big(\left[ {\FIMstep{n}{n}}^{\!\!\!-1} \right]_{\scriptscriptstyle 1:2, 1:2}\big)} \, ,%
$
we define the \gls{veb} as
$%
    {\VEB}_{,n} \!:=\! \sqrt{	\tr \big(\left[ {\FIMstep{n}{n}}^{\!\!\!-1} \right]_{\scriptscriptstyle 3:4, 3:4}\big)} \, ,%
$
we define the \gls{oeb} of the receiving agent as
$
    {\OEBRX}_{,n} \!:=\! \sqrt{ \left[ {\FIMstep{n}{n}}^{\!\!\!-1} \right]_{\scriptscriptstyle 5, 5}} \, ,%
$
we define the \gls{meb} of \gls{surface} $s$ as
$
    {\MEB{s}}_{\!,n} \!:=\! \sqrt{\tr \big(\left[ {\FIMstep{n}{n}}^{\!\!\!-1} \right]_{\scriptscriptstyle 6+2(s\!-\!1):5+2s, %
\scriptscriptstyle 6+2(s\!-\!1):5+2s}\big)} \, .
$

\def\datapath{.}
\begin{figure*}[t]
	\hspace{-32mm}%
	\setlength{\figurewidth}{0.9425\linewidth}
    \setlength{\figureheight}{0.18\linewidth}
	\input{figures/MVA-5plot.tex}
	\vspace{-6mm}
	\caption{Mapping error $\lVert \pmva{s} - \pmvaHat{s} \rVert$ for $s\in\{1\hdots 5\}$ (left to right) vs. \ac{meb} evaluated on the simulated data from~\cite{XuhongTWC2025}.}%
	\label{fig:mapping-accuracy}%
	\vspace{-1.5mm}
\end{figure*}

%% file: figures/results/synResultColorBar.tex
\definecolor{mycolor1}{rgb}{0.00000,0.50000,1.00000}%
\definecolor{amberSAE}{rgb}{1,0.49,0}

\pgfplotsset{every axis/.append style={
  label style={font=\footnotesize},
  legend style={font=\footnotesize},
  tick label style={font=\footnotesize},
}}

\pgfdeclarelayer{background layer}
\pgfdeclarelayer{foreground layer}
\pgfdeclarelayer{veryTop layer}
\pgfsetlayers{background layer,main,foreground layer,veryTop layer}

\begin{tikzpicture}
	\tikzstyle{TextNode}=[text=black,font=\normalsize];
	\tikzstyle{every node}=[font=\normalsize]
	\tikzstyle{Line} = [draw=black, dashed, line width=0.4pt];
	\tikzstyle{trueWallLinefont} = [draw=gray!90, line width=2.5pt, opacity=0.8];
	\tikzstyle{estWallLinefont} = [draw=blue, dashdotted, line width = 0.25mm, opacity=1];
	\tikzstyle{truePathfont} = [draw=black, dashed, line width = 0.1mm, opacity=0.6];
	\tikzstyle{estPathfont} = [draw=blue, dashed, line width = 1.2mm, opacity=0.6];
	\centering
	\begin{axis}[
		width=0.4\textwidth,
		height=0.05\textwidth,
		scale only axis,
		axis line style={draw=none},
		ticks=none,
		line cap = round,				
		line join = round, 				
		colormap = {whiteblack}{color(0cm)=(yellow!50);color(0.5cm)=(red!30!yellow);color(1cm)=(red!99)},
		colorbar horizontal,
		point meta min=8,
		point meta max=30,
		colorbar style={
			at={(0.04885,0.5)},anchor=north west, 
			xlabel={SNR in dB},	
			width=1.1765*\pgfkeysvalueof{/pgfplots/parent axis width},
			height=0.2*\pgfkeysvalueof{/pgfplots/parent axis height},
					line cap = round,				
		line join = round, 				
			xticklabel style={yshift=3.5*\pgfkeysvalueof{/pgfplots/height}},
			every axis x label/.style=
			{at={(ticklabel cs:0.5)},rotate=0, anchor=near ticklabel, outer sep=-10.5mm},
		},
		]

	\end{axis}

\end{tikzpicture}

%% file: figures/results/synResultPA1fig3.tex
\definecolor{mycolor1}{rgb}{0.00000,0.50000,1.00000}%
\definecolor{amberSAE}{rgb}{1,0.49,0}

\pgfdeclarelayer{background layer}
\pgfdeclarelayer{foreground layer}
\pgfdeclarelayer{veryTop layer}
\pgfsetlayers{background layer,main,foreground layer,veryTop layer}

\pgfplotsset{every axis/.append style={
  label style={font=\footnotesize},
  legend style={font=\footnotesize},
  tick label style={font=\footnotesize},
}}

\definecolor{mycolor2}{rgb}{0.00000,0.38431,0.60784}
\definecolor{mycolor3}{rgb}{0.88235,0.63922,0.00000}%
\definecolor{mycolor1}{rgb}{0.00000,0.38431,0.60784}%
\definecolor{mycolor4}{rgb}{0.52549,0.12157,0.25490}%
\definecolor{mycolor3x}{rgb}{0.00000,0.15686,0.33333}%
\definecolor{mycolor2x}{rgb}{0.00000,0.38431,0.60784}%
\definecolor{mycolor1x}{rgb}{0.88235,0.63922,0.00000}%
\definecolor{mycolor4x}{rgb}{0.72941,0.04706,0.18431}%
\definecolor{mycolor5x}{rgb}{0.52549,0.12157,0.25490}%

\begin{tikzpicture}
	\tikzstyle{TextNode}=[text=black,font=\normalsize];
	\tikzstyle{every node}=[font=\normalsize ,line cap = round, line join = round]
	\tikzstyle{Line} = [draw=black, dashed, line width=0.4pt,line cap = round, line join = round,];
	\tikzstyle{trueWallLinefont} = [draw=gray!90, line width=2.5pt, opacity=0.8,line cap = round, line join = round,];
	\tikzstyle{estWallLinefont} = [draw=blue, dashdotted, line width = 0.25mm, opacity=1,line cap = round, line join = round,];
	\tikzstyle{truePathfont} = [draw=black, dashed, line width = 0.1mm, opacity=0.6,line cap = round, line join = round,];
	\tikzstyle{estPathfont} = [draw=blue, dashed, line width = 1.2mm, opacity=0.6,line cap = round, line join = round,];
	\centering
	\begin{axis}[
		width=0.32\textwidth,
		line cap = round,				
		line join = round, 				
		scale only axis,
		xmin=-1.8,
		xmax=5.5,
		ymin=-1.5, 
		ymax=8, 
		axis equal image,				
		ylabel={$ y $ (\SI{}{\metre})},				
		xlabel={$ x $ (\SI{}{\metre})},				
		xtick={0, 2, 4},
		every axis x label/.style=
		{at={(ticklabel cs:0.53)},anchor=near ticklabel, outer sep=-1mm},
		every axis y label/.style= {at={(ticklabel cs:0.5)},rotate=90, anchor=near ticklabel, outer sep=-1mm,yshift = -1mm},
		legend style={legend columns=3,/tikz/every even column/.append style={column sep=0.2cm},legend cell align=left, at={(1.04,-0.14)}, fill=none, font=\scriptsize},
		colormap = {whiteblack}{color(0cm)=(yellow!50);color(0.5cm)=(red!30!yellow);color(1cm)=(red!99)},
		point meta min=8,
		point meta max=30,
		colorbar style={
			at={(1.05,1)}, 
			ylabel={SNR in (dB)},	
			width=0.05*\pgfkeysvalueof{/pgfplots/parent axis width},
			ytick={8, 30},
			every axis y label/.style=
			{at={(ticklabel cs:0.5)},rotate=90, anchor=near ticklabel, outer sep=-5mm},
		},
		]
		

	\begin{pgfonlayer}{foreground layer}
		\addplot [draw=cyan!90, line width=1pt, forget plot,line cap = round, line join = round] table{figures/data/estAgentTrack_step18.dat};
		\addplot [draw=cyan!90,draw opacity=0.2, line width=0.5pt, forget plot,line cap = round, line join = round] table{figures/data/estAgentTrack_step307.dat};
		
		\addplot [draw=black, densely dotted, line width=0.5pt, forget plot,line cap = round, line join = round] table{figures/data/trueAgentTrack.dat};
		
		\pgfplotsforeachungrouped \n in {1,...,5}
		{\addplot[trueWallLinefont, forget plot,line cap = round, line join = round] table{figures/data/trueWalls000\n.dat};
		}
		\pgfplotsforeachungrouped \n in {1,...,5}
		{\addplot[estWallLinefont, forget plot,line cap = round, line join = round] table{figures/data/estWalls_step18_000\n.dat};
		}
		
		\node (PA1center) at (axis cs: -0.5, 6) {}; 
		
		\pgfplotsforeachungrouped \n in {1,...,10}
		{ \addplot [mesh, point meta=\thisrow{C}, line width=1pt, opacity=0.7,line cap = round, line join = round] table [x={A}, y={B}] {figures/data/estPaths_PA1_step18_\n.dat};
		}

		\pgfplotsforeachungrouped \n in {1,...,10}
		{\addplot[truePathfont, forget plot] table{figures/data/truePaths_PA1_step18_\n.dat};
		}
		
	\node[black, font=\scriptsize,fill=white,opacity=0.75,inner sep=0.2mm, rounded corners=0.5mm] at (axis cs:3.7,6.6) {$\scriptscriptstyle n=18$};
		\node (PA2center) at (axis cs: 4.2, 1.3) {}; 
		\node[black, font=\tiny,fill=white,opacity=0.75,inner sep=0.2mm, rounded corners=0.5mm] at ([xshift=0.12cm,yshift=0.37cm]PA2center) {$ \V{p}_{\mathrm{pa}}^{(2)} $};
		\node[black, font=\tiny,fill=white,opacity=0.75,inner sep=0.2mm, rounded corners=0.5mm] at ([xshift=-0mm,yshift=4mm]PA1center) {$ \V{p}_{\mathrm{pa}}^{(1)} $};
		\input{./figures/Journal_arrays_only}
		\input{./figures/Journal_array_agent_18}
		
	\node[rectangle,inner sep=0.5pt,rounded corners=0.65mm,minimum size=0.7em,
          fill=IEEEblue,
          text=white] at (axis cs: 0, 7.5470) {\adjustbox{max width=0.65em}{$\scriptstyle1$}};

    \node[rectangle,inner sep=0.5pt,rounded corners=0.65mm,minimum size=0.7em,
          fill=IEEEred,
          text=white] at (axis cs: 4.9520, 0) {\adjustbox{max width=0.65em}{$\scriptstyle2$}};

	 \node[rectangle,inner sep=0.5pt,rounded corners=0.65mm,minimum size=0.7em,
          fill=IEEEgreen,
          text=white] at (axis cs: 0, -1) {\adjustbox{max width=0.65em}{$\scriptstyle3$}};

    \node[rectangle,inner sep=0.5pt,rounded corners=0.65mm,minimum size=0.7em,
          fill=IEEEviolett,
          text=white] at (axis cs: -1.305, 0) {\adjustbox{max width=0.65em}{$\scriptstyle4$}};

    \node[rectangle,inner sep=0.5pt,rounded corners=0.65mm,minimum size=0.7em,
          fill=IEEEyellow,
          text=white] at (axis cs: 1.75, 3.5) {\adjustbox{max width=0.65em}{$\scriptstyle5$}};

	\end{pgfonlayer}
	\end{axis}
\end{tikzpicture}

%% file: figures/results/synResultPA2fig3.tex
\definecolor{mycolor1}{rgb}{0.00000,0.50000,1.00000}%
\definecolor{amberSAE}{rgb}{1,0.49,0}

\pgfdeclarelayer{background layer}
\pgfdeclarelayer{foreground layer}
\pgfdeclarelayer{veryTop layer}
\pgfsetlayers{background layer,main,foreground layer,veryTop layer}

\pgfplotsset{every axis/.append style={
  label style={font=\footnotesize},
  legend style={font=\footnotesize},
  tick label style={font=\footnotesize},
}}

\definecolor{mycolor2}{rgb}{0.00000,0.38431,0.60784}
\definecolor{mycolor3}{rgb}{0.88235,0.63922,0.00000}%
\definecolor{mycolor1}{rgb}{0.00000,0.38431,0.60784}%
\definecolor{mycolor4}{rgb}{0.52549,0.12157,0.25490}%
\definecolor{mycolor3x}{rgb}{0.00000,0.15686,0.33333}%
\definecolor{mycolor2x}{rgb}{0.00000,0.38431,0.60784}%
\definecolor{mycolor1x}{rgb}{0.88235,0.63922,0.00000}%
\definecolor{mycolor4x}{rgb}{0.72941,0.04706,0.18431}%
\definecolor{mycolor5x}{rgb}{0.52549,0.12157,0.25490}%

\begin{tikzpicture}
	\tikzstyle{TextNode}=[text=black,font=\normalsize];
	\tikzstyle{every node}=[font=\normalsize,,line cap = round, line join = round]
	\tikzstyle{Line} = [draw=black, dashed, line width=0.4pt,line cap = round, line join = round,];
	\tikzstyle{trueWallLinefont} = [draw=gray!90, line width=2.5pt, opacity=0.8,line cap = round, line join = round,];
	\tikzstyle{estWallLinefont} = [draw=blue, dashdotted, line width = 0.25mm, opacity=1,line cap = round, line join = round,];
	\tikzstyle{truePathfont} = [draw=black, dashed, line width = 0.1mm, opacity=0.6,line cap = round, line join = round,];
	\tikzstyle{estPathfont} = [draw=blue, line width = 1.2mm, opacity=0.6,line cap = round, line join = round,];
	\centering
	\begin{axis}[
		width=0.32\textwidth,
		line cap = round,				
		line join = round, 				
		scale only axis,
		xmin=-1.8,
		xmax=5.5,
		ymin=-1.5, 
		ymax=8, 
		axis equal image,				
		xlabel={$ x $ (\SI{}{\metre})},				
		xtick={0, 2, 4},
		every axis x label/.style=
		{at={(ticklabel cs:0.53)},anchor=near ticklabel, outer sep=-1.5mm},
		every axis y label/.style= {at={(ticklabel cs:0.5)},rotate=90, anchor=near ticklabel, outer sep=-1mm},
		legend style={legend columns=3,/tikz/every even column/.append style={column sep=0.34cm},legend cell align=left, at={(0.01,0.5)}, anchor=south, fill=none, font=\scriptsize,xshift=-1mm},
		legend to name = legendFig_synResultPAs,
		colormap = {whiteblack}{color(0cm)=(yellow!50);color(0.5cm)=(red!30!yellow);color(1cm)=(red!99)},
		point meta min=8,
		point meta max=30,
		colorbar style={
			at={(1.05,1)}, 
			ylabel={SNR in (dB)},	
			width=0.05*\pgfkeysvalueof{/pgfplots/parent axis width},
			ytick={8, 30},
			every axis y label/.style=
			{at={(ticklabel cs:0.5)},rotate=90, anchor=near ticklabel, outer sep=-5mm},
		},
		]

		\addlegendimage{draw=gray!90, line width=2.5pt, opacity=0.8}\addlegendentry{\hspace{1mm}true surface}\label{pgf:true-surface}
		\addlegendimage{draw=blue, dashdotted, line width = 0.4mm, opacity=1}\addlegendentry{\hspace{1mm}est. surface}\label{pgf:est-surface}
		 
		\addlegendimage{draw=black, densely dotted, line width=0.8pt}\addlegendentry{\hspace{1mm}true agent track}\label{pgf:true-agent}
		\addlegendimage{draw=cyan, line width=2pt, opacity=0.7}\addlegendentry{\hspace{1mm}est. agent track} \label{pgf:est-agent}
		
		\addlegendimage{draw=black, dashed, line width = 0.3mm, opacity=0.9}\addlegendentry{\hspace{1mm}true paths}\label{pgf:true-paths}
		\addlegendimage{draw=red!60!yellow, line width=2pt, opacity=0.7}\addlegendentry{\hspace{1mm}est. paths}\label{pgf:est-paths}


		\begin{pgfonlayer}{foreground layer}
			\addplot [draw=cyan!90, line width=1pt, forget plot,line cap = round, line join = round] table{figures/data/estAgentTrack_step18.dat};
			\addplot [draw=cyan!90,draw opacity=0.2, line width=0.5pt, forget plot,line cap = round, line join = round] table{figures/data/estAgentTrack_step307.dat};
			
			\addplot [draw=black, densely dotted, line width=0.5pt, forget plot,line cap = round, line join = round] table{figures/data/trueAgentTrack.dat};

			\pgfplotsforeachungrouped \n in {1,...,5}
			{\addplot[trueWallLinefont, forget plot,line cap = round, line join = round] table{figures/data/trueWalls000\n.dat};
			}
			\pgfplotsforeachungrouped \n in {1,...,5}
			{\addplot[estWallLinefont, forget plot,line cap = round, line join = round] table{figures/data/estWalls_step18_000\n.dat};
			}
			
			\node (PA1center) at (axis cs: -0.5, 6) {}; 
			\node (PA2center) at (axis cs: 4.2, 1.3) {}; 
				
			\pgfplotsforeachungrouped \n in {1,...,5}
			{ \addplot [mesh, point meta=\thisrow{C}, line width=1pt, opacity=0.7,line cap = round, line join = round] table [x={A}, y={B}] {figures/data/estPaths_PA2_step18_\n.dat};
			}
			
			\pgfplotsforeachungrouped \n in {1,...,5}
			{\addplot[truePathfont, forget plot,line cap = round, line join = round] table{figures/data/truePaths_PA2_step18_\n.dat};
			}
			
			
			\node(AgentCenter) at (axis cs: 1.22, 6.81) {}; 
			\node[above,align=center,black, font=\tiny] at ([xshift=-0.8mm,yshift=0mm]AgentCenter) {$ \pos{n} $};
		\node[black, font=\scriptsize,fill=white,opacity=0.75,inner sep=0.2mm, rounded corners=0.5mm] at (axis cs:3.7,6.6) {$\scriptscriptstyle n=18$};
					\node[black, font=\tiny,fill=white,opacity=0.75,inner sep=0.2mm, rounded corners=0.5mm] at ([xshift=0.12cm,yshift=0.37cm]PA2center) {$ \V{p}_{\mathrm{pa}}^{(2)} $};
		\node[black, font=\tiny,fill=white,opacity=0.75,inner sep=0.2mm, rounded corners=0.5mm] at ([xshift=-0mm,yshift=4mm]PA1center) {$ \V{p}_{\mathrm{pa}}^{(1)} $};
						\input{./figures/Journal_arrays_only}
						\input{./figures/Journal_array_agent_18}

	\node[rectangle,inner sep=0.5pt,rounded corners=0.65mm,minimum size=0.7em,
          fill=IEEEblue,
          text=white] at (axis cs: 0, 7.5470) {\adjustbox{max width=0.65em}{$\scriptstyle1$}};

    \node[rectangle,inner sep=0.5pt,rounded corners=0.65mm,minimum size=0.7em,
          fill=IEEEred,
          text=white] at (axis cs: 4.9520, 0) {\adjustbox{max width=0.65em}{$\scriptstyle2$}};

	 \node[rectangle,inner sep=0.5pt,rounded corners=0.65mm,minimum size=0.7em,
          fill=IEEEgreen,
          text=white] at (axis cs: 0, -1) {\adjustbox{max width=0.65em}{$\scriptstyle3$}};

    \node[rectangle,inner sep=0.5pt,rounded corners=0.65mm,minimum size=0.7em,
          fill=IEEEviolett,
          text=white] at (axis cs: -1.305, 0) {\adjustbox{max width=0.65em}{$\scriptstyle4$}};

    \node[rectangle,inner sep=0.5pt,rounded corners=0.65mm,minimum size=0.7em,
          fill=IEEEyellow,
          text=white] at (axis cs: 1.75, 3.5) {\adjustbox{max width=0.65em}{$\scriptstyle5$}};

	\addplot [color=black, only marks, mark=*, mark options={solid, IEEEblue},  mark size=1pt, forget plot]
	 	 table[row sep=crcr]{%
		0	0\\
		};
	\node[below, inner sep=0]
		at (axis cs:0,-0.15) {$\scriptstyle\bm{0}$};
		
\addplot[-stealth, color=black, line width=0.601pt,%
{-Stealth[inset=0pt, scale=1.05, angle'=25]}
]
  table[row sep=crcr]{%
0.15	0\\
0.8	0\\
};
\node[centered, align=center, inner sep=0]
at (axis cs:-0.2,0.8) {$\scriptstyle y$};

\addplot[-stealth, color=black, line width=0.601pt,%
{-Stealth[inset=0pt, scale=1.05, angle'=25]}
]
  table[row sep=crcr]{%
0	0.15\\
0	0.8\\
};
\node[centered, align=center, inner sep=0]
at (axis cs:0.8,-0.2) {$\scriptstyle x$};

		\end{pgfonlayer}

	\end{axis}

\end{tikzpicture}

%% file: figures/MVA-5plot.tex
%
%

\pgfplotsset{every axis/.append style={
  label style={font=\footnotesize},
  legend style={font=\scriptsize},
  tick label style={font=\footnotesize},
}}

\newcommand{\lblMVAest}[1]{RMSE}
\newcommand{\lblMVAmeb}[1]{MEB}
\newcommand{\lblMVAuncertainty}[1]{$\interval{80\%}$}

\begin{tikzpicture}

\begin{axis}[%
axis line style = thick,	
width=0.195\figurewidth,
height=\figureheight,
at={(0\figurewidth,0\figureheight)},
scale only axis,
xmin=1,
xmax=307,
xlabel style={yshift=2mm},
xlabel={Step $n$},
ymin=0,
ymax=100,
ylabel style={yshift=-4mm},
ylabel={$\lVert \pmva{s} - \pmvaHat{s} \rVert$ in \SI{}{\centi\metre}},
legend style={legend cell align=left, align=left, draw=white!15!black}
]
\node[below, align=center,%
    rectangle,inner sep=0.5pt,rounded corners=0.65mm,minimum size=0.7em,
          fill=IEEEblue,
          text=white] at (axis cs:40,98) {\adjustbox{max width=0.65em}{$\scriptstyle1$}};

\addplot[blue, line width = 0.15mm, opacity=1, densely dashed, line width=\LWestimates] table{./figures/datFiles/meanPositionError_MVA3_snr0001.dat};\label{pgf:rmseMVA}\addlegendentry{\lblMVAest{3}}

\addplot[blue, opacity=0.7, line width=\LWbound] table{./figures/datFiles/MPEBMapping_MVA3_snr0001.dat};	%
\label{pgf:MEB}%
\addlegendentry{\lblMVAmeb{3}}

\addplot[name path=down,color=blue!30, opacity=0,forget plot] table{./figures/datFiles/meanPositionErrorVarLower_MVA3_snr0001.dat};
\addplot[name path=up,color=blue!30, opacity=0,forget plot] table{./figures/datFiles/meanPositionErrorVarUpper_MVA3_snr0001.dat};
\addplot[blue!30,fill opacity=0.4] fill between[of=down and up];%
\label{pgf:sigmaMVA}%
\addlegendentry{\lblMVAuncertainty{3}}
\end{axis}

\begin{axis}[%
axis line style = thick,	
width=0.195\figurewidth,
height=\figureheight,
at={(0.2\figurewidth,0\figureheight)},
scale only axis,
xmin=1,
xmax=307,
xlabel style={yshift=2mm},
xlabel={Step $n$},
ymin=0,
ymax=100,
yticklabel=\empty,                      
legend style={legend cell align=left, align=left, draw=white!15!black}
]
\node[below, align=center,%
    rectangle,inner sep=0.5pt,rounded corners=0.65mm,minimum size=0.7em,
          fill=IEEEred,
          text=white] at (axis cs:40,98) {\adjustbox{max width=0.65em}{$\scriptstyle2$}};

\addplot[red, line width = 0.15mm, opacity=1, densely dashed, line width=\LWestimates] table{./figures/datFiles/meanPositionError_MVA2_snr0001.dat};\addlegendentry{\lblMVAest{2}}

\addplot[red, opacity=0.7, line width=\LWbound] table{./figures/datFiles/MPEBMapping_MVA2_snr0001.dat};	\addlegendentry{\lblMVAmeb{2}}

\addplot[name path=down,color=red!30, opacity=0,forget plot] table{./figures/datFiles/meanPositionErrorVarLower_MVA2_snr0001.dat};
\addplot[name path=up,color=red!30, opacity=0,forget plot] table{./figures/datFiles/meanPositionErrorVarUpper_MVA2_snr0001.dat};
\addplot[red!30,fill opacity=0.4,draw=none] fill between[of=down and up];\addlegendentry{\lblMVAuncertainty{2}}
\end{axis}

\begin{axis}[%
axis line style = thick,	
width=0.195\figurewidth,
height=\figureheight,
at={(0.4\figurewidth,0\figureheight)},
scale only axis,
xmin=1,
xmax=307,
xlabel style={yshift=2mm},
xlabel={Step $n$},
ymin=0,
ymax=100,
yticklabel=\empty,                      
legend style={legend cell align=left, align=left, draw=white!15!black}
]
\node[below, align=center,%
    rectangle,inner sep=0.5pt,rounded corners=0.65mm,minimum size=0.7em,
          fill=IEEEgreen,
          text=white] at (axis cs:40,98) {\adjustbox{max width=0.65em}{$\scriptstyle3$}};

\addplot[IEEEgreen, line width = 0.15mm, opacity=1, densely dashed, line width=\LWestimates] table{./figures/datFiles/meanPositionError_MVA1_snr0001.dat};\addlegendentry{\lblMVAest{1}}

\addplot[IEEEgreen, opacity=0.7, line width=\LWbound] table{./figures/datFiles/MPEBMapping_MVA1_snr0001.dat};	\addlegendentry{\lblMVAmeb{1}}

\addplot[name path=down,color=IEEEgreen!30, opacity=0,forget plot] table{./figures/datFiles/meanPositionErrorVarLower_MVA1_snr0001.dat};
\addplot[name path=up,color=IEEEgreen!30, opacity=0,forget plot] table{./figures/datFiles/meanPositionErrorVarUpper_MVA1_snr0001.dat};
\addplot[IEEEgreen!30,fill opacity=0.4,draw=none] fill between[of=down and up];\addlegendentry{\lblMVAuncertainty{1}}
\end{axis}

\begin{axis}[%
axis line style = thick,	
width=0.195\figurewidth,
height=\figureheight,
at={(0.599\figurewidth,0\figureheight)},
scale only axis,
xmin=1,
xmax=307,
xlabel style={yshift=2mm},
xlabel={Step $n$},
ymin=0,
ymax=100,
yticklabel=\empty,                      
legend style={legend cell align=left, align=left, draw=white!15!black}
]
\node[below, align=center,%
    rectangle,inner sep=0.5pt,rounded corners=0.65mm,minimum size=0.7em,
          fill=IEEEviolett,
          text=white] at (axis cs:40,98) {\adjustbox{max width=0.65em}{$\scriptstyle4$}};

\addplot[IEEEviolett, line width = 0.15mm, opacity=1, densely dashed, line width=\LWestimates] table{./figures/datFiles/meanPositionError_MVA4_snr0001.dat};\addlegendentry{\lblMVAest{4}}

\addplot[IEEEviolett, opacity=0.7, line width=\LWbound] table{./figures/datFiles/MPEBMapping_MVA4_snr0001.dat};	\addlegendentry{\lblMVAmeb{4}}

\addplot[name path=down,color=IEEEviolett!30, opacity=0,forget plot] table{./figures/datFiles/meanPositionErrorVarLower_MVA4_snr0001.dat};
\addplot[name path=up,color=IEEEviolett!30, opacity=0,forget plot] table{./figures/datFiles/meanPositionErrorVarUpper_MVA4_snr0001.dat};
\addplot[IEEEviolett!30,fill opacity=0.4,draw=none] fill between[of=down and up];\addlegendentry{\lblMVAuncertainty{4}}
\end{axis}

\begin{axis}[%
axis line style = thick,	
width=0.195\figurewidth,
height=\figureheight,
at={(0.799\figurewidth,0\figureheight)},
scale only axis,
xmin=1,
xmax=307,
xlabel style={yshift=2mm},
xlabel={Step $n$},
ymin=0,
ymax=100,
yticklabel=\empty,                      
legend style={legend cell align=left, align=left, draw=white!15!black}
]

\node[below, align=center,%
    rectangle,inner sep=0.5pt,rounded corners=0.65mm,minimum size=0.7em,
          fill=IEEEyellow,
          text=white] at (axis cs:40,98) {\adjustbox{max width=0.65em}{$\scriptstyle5$}};

\addplot[mycolor5, line width = 0.15mm, opacity=1, densely dashed, line width=\LWestimates] table{./figures/datFiles/meanPositionError_MVA5_snr0001.dat};\addlegendentry{\lblMVAest{5}}

\addplot[mycolor5, opacity=0.7, line width=\LWbound] table{./figures/datFiles/MPEBMapping_MVA5_snr0001.dat};	\addlegendentry{\lblMVAmeb{5}}

\addplot[name path=down,color=mycolor5!30, opacity=0,forget plot] table{./figures/datFiles/meanPositionErrorVarLower_MVA5_snr0001.dat};
\addplot[name path=up,color=mycolor5!30, opacity=0,forget plot] table{./figures/datFiles/meanPositionErrorVarUpper_MVA5_snr0001.dat};
\addplot[mycolor5!30,fill opacity=0.4,draw=none] fill between[of=down and up];\addlegendentry{\lblMVAuncertainty{5}}

\end{axis}

\end{tikzpicture}%

%% file: InputFiles/ProblemFormulation.tex
The goal in \ac{dmimo} SLAM is to estimate the agent- and map (surface) states which we do by means of the \ac{mmse} estimator\footnote{In the scenario (see Fig.\,\ref{fig:synResultTwoPAs}) we assume that all surfaces $s\!\in\!\setWalls$ exist at all times in the sense that at least one path $\MVApair{s}{s'}$ is visible at all times $n$.}~\cite{Kay_EstimationTheory}
\begin{align}	
	\stateHat{n} &:= \mathbb{E}(\RVstate{n}|\RVobservationMatrix{1:n}\!=\!\observationMatrix{1:n}) \triangleq \int \state{n} f(\state{n}|\observationMatrix{1:n} )\mathrm{d} \state{n}\,,
	\label{eq:MMSE_x} \\
	\pmvaHat{s} &\!:=\! \mathbb{E}(\RVpmva{s}|\RVobservationMatrix{1:n}\!=\!\observationMatrix{1:n})\! \triangleq\!\! \int \pmva{s} f(\pmva{s} |\observationMatrix{1:n} )\mathrm{d} \pmva{s}\,,
	\nonumber 
\end{align}
i.e., the expectations under the \textit{marginal} posterior \acp{pdf} $f(\RVstate{n}|\RVobservationMatrix{1:n}\!\!=\!\!\observationMatrix{1:n})$ and $f(\RVpmva{s}|\RVobservationMatrix{1:n}\!\!=\!\!\observationMatrix{1:n})$.
We obtain these marginal posterior \acp{pdf} by applying the \ac{spa}~\cite[Sec.\,8.4.4]{Bishop} to the factor graph in~\cite{XuhongTWC2025} which is a graphical representation the \textit{joint} posterior \ac{pdf} of all random variables of our statistical model conditional on the observations $\observationMatrix{1:n}$ made over all time steps $1\!:\!n$ by all anchors $j$.
The detection problem, i.e., inference of the existences $\rv{r}_{ss',n}^{(j)}$,
and the data association problem, i.e., inference of the correct association between measurements $m$ and \acp{mpc} $\MVApair{s}{s'}$, complicate the joint estimation problem.
The latter in particular, as it introduces cycles in the factor graph.
While the \ac{spa} would result in \textit{exact} inference in tree-structured graphs, \textit{approximate} inference~\cite[Sec.\,V]{FG_SPA_TIT2001} is still possible by iteratively applying the \gls{spa} resulting in loopy \gls{bp}~\cite{Frey97loopyBP}.
Our method detailed in~\cite{XuhongTWC2025} is 
an extension of~\cite{Erik_SLAM_TWC2019, XuhongTWC2022, LeiVenTeaMey:TSP2023} and leverages the instance of loopy \gls{bp} for scalable data association from~\cite{Florian_Proceeding2018}.

%% file: InputFiles/Results.tex
	
	\tikzset{
  		every picture/.append style={
    		line cap=round,
    		line join=round}
	}
	\newcommand{\MCruns}{ {100} }							
\newcommand{\OSPAcutoff}{ {\SI{5}{\metre}} }			
\newcommand{\LWbound}{1pt}
\newcommand{\LWestimates}{0.5pt}
\definecolor{mycolor1}{rgb}{0.00000,0.38431,0.60784}%
\definecolor{mycolor2}{rgb}{0.72941,0.04706,0.18431}%
\definecolor{mycolor3}{rgb}{0.00000,0.51765,0.23922}%
\definecolor{mycolor4}{rgb}{0.59608,0.11373,0.59216}%
\definecolor{mycolor5}{rgb}{0.88235,0.63922,0.00000}%
\colorlet{mycolor1}{IEEEblue}
\colorlet{mycolor2}{IEEEred}
\colorlet{mycolor3}{IEEEgreen}
\colorlet{mycolor4}{IEEEviolett}
\colorlet{mycolor5}{IEEEorange}
%

\def\datapath{.}
\begin{figure}[t]
	\setlength{\figurewidth}{0.95\linewidth}
    \setlength{\figureheight}{0.3\linewidth}
	\hspace{-25mm}\input{figures/RMSE-agent-pos.tex}%
	\vspace{-2mm}%
	\hspace{-25.0mm}\input{figures/RMSE-agent-orient.tex}%
	\vspace{-2mm}
	\caption{Performance of the algorithm in~\cite{XuhongTWC2025} vs. \ac{pcrlb}: a) RMSE of the agent position and b) RMSE of the agent orientation versus \ac{pcrlb} evaluated over time on simulated data.}%
	\label{fig:RMSE-agent}%
	\vspace{-1.5mm}
\end{figure}

{\slshape Experiment.} We simulate synthetic measurements in the scenario depicted in Fig.\,\ref{fig:synResultTwoPAs} with $J\!=\!2$ \acp{pa}, $S\!=\!5$ \acp{surface}, and the mobile agent moving on a trajectory $\{\pos{n}\}_{n=1}^{N}$ with $N\!=\!307$ steps.
We run the algorithm from~\cite{XuhongTWC2025} on the 
simulated measurements and compare its estimation performance against the \ac{pcrlb} derived in Sec.\,\ref{sec:PCRLB}.

{\slshape Results.} %
Representing localization results, Fig.\,\ref{fig:synResultTwoPAs} shows the estimated agent track~\lineref{pgf:est-agent} vs. the true agent track~\lineref{pgf:true-agent}. 
The mapping results show estimated \acp{surface}~\lineref{pgf:est-surface} vs. the true surface positions and orientations~\lineref{pgf:true-surface}. 
Our current algorithm assumes walls with infinite extents. 
Future work will also estimate wall extents. 
Ground truth paths~\lineref{pgf:true-paths} for all visible components $\MVApair{s}{s'}\! \in\!\setSDBt$ along with estimated paths~\lineref{pgf:est-paths} for \ac{pa} $j\!=\!1$ and \ac{pa} $j\!=\!2$ are shown in Fig.\,\ref{fig:synResultTwoPAs}\,a) and b), respectively.
Some paths such as the \ac{los} path of \ac{pa} $2$ are obstructed by the center surface \wallrefE\, leading to $\rv{r}_{00,n}^{(2)}\!=\!0$ at time $n\!=\!18$.

Figures~\ref{fig:mapping-accuracy} and~\ref{fig:RMSE-agent} show the estimation errors computed over a \ac{mc} analysis with $\MCruns$ runs of different random measurement realizations and algorithm initializations. 
The estimation errors are compared against the \ac{pcrlb}.
Fig.\,\ref{fig:mapping-accuracy} shows the \acp{rmse}~\lineref{pgf:rmseMVA} 
of the \ac{surface} estimates $\pmvaHat{s}$ for $S\!=\!5$ surfaces \wallrefA\,--\,\wallrefE~that are marked in Fig.\,\ref{fig:synResultTwoPAs}.
We augment all figures showing estimation errors with central $p=80\%$ sample-quantile intervals $\interval{p}:=[\hat{L}(p),\hat{U}(p)]$~\lineref{pgf:sigmaMVA} for which one new random error sample\footnote{Computed through the expression annotating the respective vertical axis.} $|e|$ is located within the interval with a probability of $p$, i.e., $p \approx \mathbb{P}\big(\hat{L}(p) \leq |\rv{e}| \leq \hat{U}(p)\big)$. 
Based on the \ac{mc} analysis, the interval is an ensemble-realization of the limits $L(p)\!=\!q_{\frac{1-p}{2}}$ and $U(p)\!=\!q_{\frac{1+p}{2}}$ computed through population quantiles ${q}_\alpha\!:=\! \inf\left\{x~|~\mathbb{P}(|\rv{e}|\leq x)\geq\alpha\right\}$.
Two diverged runs were discarded. 
In addition, we removed all outliers with values more than $1.5$ inter-quartile ranges above the upper quartile or below the lower quartile leading to a rejection rate of $\SI{3.3}{\percent}$.
\ac{surface} estimates and ground truth were associated using the Hungarian method~\cite{Kuhn1955Hungarian}, missed detections and false alarms not evaluated.
The algorithm from~\cite{XuhongTWC2025} asymptotically approaches the \acp{meb} 
for all surfaces. 
Note that all single-bounce paths at \ac{surface} 
\wallrefC~are \textit{not} visible 
($\rv{r}_{33,n}^{(j)}\!=\!0~\forall j$)
until $n\!>\!125$~(cf.\,Fig.\,\ref{fig:synResultTwoPAs}), demonstrating both the potential of using double-bounce paths for mapping as well as the efficacy of the algorithm.

Fig.\,\ref{fig:RMSE-agent}\,a) shows the \ac{rmse} of the agent position estimates which is rather low from the first time steps and gets very close to the \ac{peb} which is ${\PEB}_{,n}\!<\!\SI{1}{\centi\metre}$ for all time steps $n$, highlighting the high accuracy achievable with \ac{mpslam} in MIMO systems.
In Fig.\,\ref{fig:RMSE-agent}\,b), we see that the \ac{rmse} of the agent orientation estimates w.r.t. the ground truth likewise approaches the~\ac{oeb} after the first few time steps. 
The estimation error on the agent orientation in MIMO systems is generally low~\cite{Guerra18OEB,DeutschmannCISA2025}.

%% file: figures/RMSE-agent-pos.tex
%
%
\definecolor{mycolor1}{rgb}{0.00000,0.38431,0.60784}%

\pgfplotsset{every axis/.append style={
  label style={font=\footnotesize},
  legend style={font=\scriptsize},
  tick label style={font=\footnotesize},
}}

\begin{tikzpicture}

\begin{axis}[%
axis line style = thick,	
width=0.951\figurewidth,
height=\figureheight,
at={(0\figurewidth,0\figureheight)},
scale only axis,
unbounded coords=jump,
xmin=2,
xmax=307,
ylabel={$\lVert \V{p}_{n} - \hat{\V{p}}_{n} \rVert$ in \SI{}{\centi\metre}},	
ylabel style={yshift=-7mm},	
ymin=0,
ymax=8,
ytick distance=1,
xticklabel=\empty,                      
legend style={legend cell align=left, align=left, draw=white!15!black}
]
\addplot [blue, line width = \LWestimates, opacity=1] table {./figures/datFiles/meanRMSEagentPosErrorNew_snr0001.dat};
\addlegendentry{RMSE}

\addplot[name path=down,color=blue, opacity=0, forget plot] table{./figures/datFiles/stdAgentPositionLowerNewSLAM_snr0001.dat};
\addplot[name path=up,color=blue!30, opacity=0, forget plot] table{./figures/datFiles/stdAgentPositionUpperNewSLAM_snr0001.dat};
\addplot[blue!30,fill opacity=0.4] fill between[of=down and up];
\addlegendentry{$\interval{80\%}$}

\addplot[red, line width = \LWbound,  opacity=0.9, mark options={solid}] table {figures/datFiles/meanPEB_AgentPosition_snr0001.dat};
\addlegendentry{PEB}

\node[right, align=left,font={\footnotesize},fill=white,
opacity=0.9,inner sep=0.75mm, xshift=-0.1mm, yshift=2mm, draw, line width = 0.8pt] at %
(axis cs:2,0){a)};

\end{axis}
\end{tikzpicture}%

%% file: figures/RMSE-agent-orient.tex
%
%
\definecolor{mycolor1}{rgb}{0.72941,0.04706,0.18431}%

\pgfplotsset{every axis/.append style={
  label style={font=\footnotesize},
  legend style={font=\scriptsize},
  tick label style={font=\footnotesize},
}}

\renewcommand{\LWbound}{1pt}
\renewcommand{\LWestimates}{0.5pt}

\begin{tikzpicture}

\begin{axis}[%
axis line style = thick,	
width=0.951\figurewidth,
height=\figureheight,
at={(0\figurewidth,0\figureheight)},
scale only axis,
unbounded coords=jump,
xmin=1,
xmax=307,
ymin=0,
ymax=4,
ylabel={$| {{\Delta\varphi}}_{n} - \widehat{{\Delta\varphi}}_{n} |$ in \SI{}{\degree}},	
ylabel style={yshift=-7.4mm},	
xlabel={Step $n$},
xlabel style={yshift=2.5mm},	
legend style={legend cell align=left, align=left, draw=white!15!black}
]
\addplot [blue, line width = \LWestimates, opacity=1] table {./figures/datFiles/meanRMSEagentOriErrorNew_snr0001.dat};
\addlegendentry{RMSE}

\addplot[name path=down,color=blue, opacity=0, forget plot] table{./figures/datFiles/stdAgentOrientationLowerNewSLAM_snr0001.dat};
\addplot[name path=up,color=blue!30, opacity=0, forget plot] table{./figures/datFiles/stdAgentOrientationUpperNewSLAM_snr0001.dat};
\addplot[blue!30,fill opacity=0.4] fill between[of=down and up];
\addlegendentry{$\interval{80\%}$}

\addplot[red, line width = \LWbound,  opacity=0.9, mark options={solid}] table {figures/datFiles/meanPEB_AgentOri_snr0001.dat};
\addlegendentry{OEB}

\node[right, align=left,font={\footnotesize},fill=white,
opacity=0.9,inner sep=0.75mm, xshift=-0.3mm, yshift=2mm, draw, line width = 0.8pt] at %
(axis cs:2,0){b)};
\end{axis}

\end{tikzpicture}%

%% file: InputFiles/appendix.tex
\section{Azimuth Rotation Matrix}\label{eq:azRotM}
We define the azimuth rotation as the mapping
$\bm{R} \!:\! (-\pi,\pi] \!\to\! SO(2),~ \phi\! \mapsto \!\bm{R}(\phi)$
where the 
rotation matrix
\begin{align}\label{eq:rotM}
	\bm{R}(\phi) =
	\begin{bmatrix}
	\cos(\phi) &-\sin(\phi) \\
     \sin(\phi) & ~\cos(\phi) 
	\end{bmatrix}
\end{align}
models counterclockwise rotations of \acp{pa} or agent 
and
\begin{align}\label{eq:rotMdot}
	\dot{\bm{R}}(\phi) := \frac{\diff \bm{R}(\phi) }{\diff \phi} =
	\begin{bmatrix}
	-\sin(\phi) & -\cos(\phi) \\
      \cos(\phi) & -\sin(\phi)
	\end{bmatrix}
\end{align}
is its derivative w.r.t. the rotation angle $\phi$.


\section{Derivation of Jacobian Matrices}\label{sec:app-global-FIM}

%
As a result of the chain rule, the Jacobian submatrices in~\eqref{eq:jacobian-main} compute as the product of more fundamental Jacobian building blocks, which we derive first:

Abbreviating $\rx:=[\rangep{n}{j}{s}{s'}]_{\scriptscriptstyle 1}$ and $\ry:=[\rangep{n}{j}{s}{s'}]_{\scriptscriptstyle 2}$ 
for notational brevity, the mapping of Fisher information in \textit{local} spherical \gls{pa} coordinates to \textit{local} Cartesian \gls{pa} coordinates 
in azimuth $\azx{j} = \arctantwo (\ry,\rx)$ is~(cf.\,\cite[eq.\,(S23)]{ Fascista25RadioStripes}) 
\begin{align}
     { \frac{\partial \azx{j} }{\partial\rangep{n}{j}{s}{s'} }} = 
    \frac{1}{\rx^2 + \ry^2}
    \begin{bmatrix}
        -\ry \\
        \rx \\
    \end{bmatrix} 
    \quad \in \realset{2}{1} \,.
\end{align}
Analogously, abbreviating $\rxTX:=[\rangepTX{n}{j}{s}{s'}]_{\scriptscriptstyle 1}$ and $\ryTX:=[\rangepTX{n}{j}{s}{s'}]_{\scriptscriptstyle 2}$, the mapping of Fisher information in \textit{local} spherical \gls{pm} coordinates to \textit{local} Cartesian \gls{pm} coordinates 
in azimuth $\azxAoA{j} = \arctantwo (\ryTX,\rxTX)$ is 
\begin{align}
     { \frac{\partial \azxAoA{j} }{\partial\rangepTX{n}{j}{s}{s'} }} = 
     \frac{1}{\rxTX^2 + \ryTX^2}
    \begin{bmatrix}
        -\ryTX \\
        \rxTX 
    \end{bmatrix}  
    \quad \in \realset{2}{1} \,.
\end{align}
The mapping of Fisher information in \textit{local} spherical \gls{pa} coordinates to \textit{local} Cartesian \gls{pa} coordinates in distance $\delayx{j} =  {\lVert \rangep{n}{j}{s}{s'} \rVert} $ from~\eqref{eq:DB-dist} is 
\begin{align}
     { \frac{\partial \delayx{j} }{\partial\rangep{n}{j}{s}{s'} }} = \frac{\rangep{n}{j}{s}{s'}}{ \lVert \rangep{n}{j}{s}{s'} \rVert} \quad \in \realset{2}{1} \,.
\end{align}
The mapping from \textit{local} Cartesian \gls{pa} coordinates to \textit{global} Cartesian coordinates is\footnote{$\rotM^{-1} \triangleq \rotM^\trp$ due to the orthogonality of rotation matrices.}
\begin{align}
     { \frac{\partial\rangep{n}{j}{s}{s'}\!^\trp}{\partial\ranget{n}{j}{s}{s'}} } = \rotM \quad 
    \quad \in SO(2) \,,
\end{align}
and the mapping from \textit{local} Cartesian \gls{pm} coordinates to \textit{global} Cartesian coordinates is
\begin{align}
     { \frac{\partial\rangepTX{n}{j}{s}{s'}\!^\trp}{\partial\range{n}{j}{s}{s'}} } = -\rotMn \quad 
    \quad \in SO(2) \,.
\end{align}
Fisher information about the vector $\ranget{n}{j}{s}{s'}$, pointing from \gls{pa} $j$ to the \gls{vm}, maps to the agent position $\pos{n}$ via
\begin{align}\label{eq:Jacob-ranget-pos}
     { \frac{\partial\ranget{n}{j}{s}{s'}\!^\trp}{\partial\pos{n}} } = 
    \frac{\partial\range{n}{j}{s}{s'}\!^\trp}{\partial\pos{n}} \frac{\partial\ranget{n}{j}{s}{s'}\!^\trp}{\partial\range{n}{j}{s}{s'}} = \house{s'}\,\house{s}
    \quad \in \realset{2}{2} \,,
\end{align}
using $(\house{s}\,\house{s'})^\trp = \house{s'}\,\house{s}$ and knowing that information about $\range{n}{j}{s}{s'}$, pointing from \gls{va} $j$ to the agent, maps to agent position $\pos{n}$ simply through $\partial\range{n}{j}{s}{s'}\!^\trp/\partial\pos{n} = \eye{}$.

These fundamental Jacobian building blocks will reappear in the submatrix blocks of the Jacobians $\jacobgn{j}$ in~\eqref{eq:jacobian-main}, which we define next.

\paragraph{Positioning submatrices}       
Each component $\ncomponent$ impinging at anchor $j$ leads to a $(2\times1)$ Jacobian submatrix mapping Fisher information to agent position $\pos{n}$ via
\vspace*{-2mm}
\begin{align}
    \left[\jacobP{\azimuth} \right]_{\scriptscriptstyle :,\ncomponent}
    &= 
    \frac{\partial\ranget{n}{j}{s}{s'}\!^\trp}{\partial\pos{n}}
    \frac{\partial\rangep{n}{j}{s}{s'}\!^\trp}{\partial\ranget{n}{j}{s}{s'}}
     \frac{\partial \azx{j} }{\partial\rangep{n}{j}{s}{s'} } \nonumber\\
    &= 
    \house{s'}\,\house{s} ~ \rotM ~ 
    \frac{1}{\rx^2 + \ry^2}
    \begin{bmatrix}
        -\ry \\
        \rx \\
    \end{bmatrix} \quad \in \realset{2}{1} \, ,  \\
    \left[\jacobP{\azimuthAoA} \right]_{\scriptscriptstyle :,\ncomponent}
    &= 
    \frac{\partial\range{n}{j}{s}{s'}\!^\trp}{\partial\pos{n}}
    \frac{\partial\rangepTX{n}{j}{s}{s'}\!^\trp}{\partial\range{n}{j}{s}{s'}}
    \frac{\partial \azxAoA{j} }{\partial\rangepTX{n}{j}{s}{s'} }  \nonumber\\
    &= 
    \eye{}~ \left(-\rotMn\right) ~ 
     \frac{1}{\rxTX^2 + \ryTX^2}
    \begin{bmatrix}
        -\ryTX \\
        \rxTX \\
    \end{bmatrix} \quad \in \realset{2}{1} \, ,  \\
    \left[\jacobP{\delay} \right]_{\scriptscriptstyle :,\ncomponent}
    &= 
    \frac{\partial\ranget{n}{j}{s}{s'}\!^\trp}{\partial\pos{n}}
    \frac{\partial\rangep{n}{j}{s}{s'}\!^\trp}{\partial\ranget{n}{j}{s}{s'}}
    \frac{\partial \delayx{j} }{\partial\rangep{n}{j}{s}{s'} } \nonumber \\
    &= 
    \house{s'}\,\house{s} ~ \rotM ~ 
    \frac{\rangep{n}{j}{s}{s'}}{  ~ \lVert \rangep{n}{j}{s}{s'} \rVert} \quad \in \realset{2}{1}     \, ,
\end{align}
from \gls{aod}, \gls{aoa}, and distance, respectively.

\paragraph{Orientation submatrices}           

Only \gls{aoa} maps to the agent orientation $\Delta\varphi_{n}$ via $\jacobO{\azimuthAoA}$ through
\begin{align}\label{eq:jacobO}
    \left[ \jacobO{\azimuthAoA} \right]_{\scriptscriptstyle :,\ncomponent} 
    &=     
    \frac{\partial\rangepTX{n}{j}{s}{s'}\!^\trp}{\partial\Delta\varphi_{n}}
    \frac{\partial \azxAoA{j} }{\partial\rangepTX{n}{j}{s}{s'} } 
    \nonumber \\
    &= 
    \jacobOsubTX ~ 
     \frac{1}{\rxTX^2 + \ryTX^2}
    \begin{bmatrix}
        -\ryTX \\
        \rxTX \\
    \end{bmatrix}  \quad \in \realset{1}{1}
\end{align}
with the Jacobian matrix
\begin{align}\label{eq:jacobOsubTX}
        \jacobOsubTX :=  { \frac{\partial\rangepTX{n}{j}{s}{s'}\!^\trp}{\partial\Delta\varphi_{n}}} 
    = 
    -
        \left.\range{n}{j}{s}{s'}\right.^\trp \rotMdot{n}(\Delta\varphi_{n}) 
    \quad \in \realset{1}{2}
\end{align}
mapping from vector $\rangepTX{n}{j}{s}{s'}$ in local Cartesian agent coordinates to the agent orientation $\Delta\varphi_{n}$.
The derivative of the rotation matrix $\rotMdot{n}\!:=\!\dot{\bm{R}}(\Delta\varphi_{n})\!\in\!\realset{2}{2}$ is defined in $\eqref{eq:rotMdot}$.
Note that expanding $\range{n}{j}{s}{s'} \!=\! -\rotMn \rangepRX{n}{j}{s}{s'}   $ 
in~\eqref{eq:jacobOsubTX} and reinserting $\jacobOsubTX$ in~\eqref{eq:jacobO}, it is easy to show that $\left[ \jacobO{\azimuthAoA} \right]_{\scriptscriptstyle :,\ncomponent} \triangleq -1 ~ \forall \ncomponent$ in \ac{2d}.


\paragraph{Mapping submatrices}           
For mapping, we have derived three more fundamental Jacobian blocks that map Fisher information in $\range{n}{j}{s}{s'}$ (pointing from a \gls{va} at $\posVA{j}{s}{s'}$ to the agent at $\pos{n}$) to \gls{surface} position $\pmva{s}$ 
via 
\begin{align}
    & {  \frac{\partial {\range{n}{j}{s}{s'}}^\trp }{\partial\pmva{s} }  } =: 
    \jacobMVAblock{s}{s'}  
    \\ &= 
    \begin{cases}
        2 \frac{\posAnchor{j} {\pmva{s}}^\trp}{\lVert\pmva{s}\rVert^2} \!+ \!
        2 \frac{{\posAnchor{j}}^\trp \pmva{s}}{\lVert\pmva{s}\rVert^2} \house{s} \!-\!
        \eye{} =:\jacobMVAsb 
        & \text{SB at $s$} 
        \\
        \jacobMVAsb
          ~\house{s'}
        & \text{DB $\MVApair{s}{s'}$} 
        \\
        2 \frac{\posVA{j}{s'}{s'} {\pmva{s}}^\trp}{\lVert\pmva{s}\rVert^2} \!+ \!
        2 \frac{{\posVA{j}{s'}{s'}}^\trp \pmva{s}}{\lVert\pmva{s}\rVert^2} \house{s} \! - \!
        \eye{} 
        & \text{DB $\MVApair{s'\!}{s}$} \\
    \end{cases} \nonumber
\end{align}
where we discriminate between single and double-bounce paths.
Because $\ranget{n}{j}{s}{s'}$ depends on $\pmva{s}$ through both
$\range{n}{j}{s}{s'}$ and $\house{s}$, we evaluate the Jacobian blocks
$\nicefrac{\partial\ranget{n}{j}{s}{s'}\!^\trp}{\partial\pmva{s}}$ as
\begin{equation*}
\begin{adjustbox}{max width=\columnwidth}
$\displaystyle
\begin{aligned}
    &\jacobMVAblockG{s}{s'}
    \!\coloneqq\!
    \frac{\partial\ranget{n}{j}{s}{s'}\!^{\!\trp}}{\partial 
    \pmva{s}}
\\
  &\!=\! 
    \begin{cases}
    \!\Bigg(\!
    \jacobMVAsb
    \!-\!
    2
    \bigg(\!
        \frac{
            {\range{n}{j}{s}{s}}^{\trp}
            \pmva{s}
        }{
            \lVert \pmva{s} \rVert^2
        }
        \eye{}
        \!-\!
        \frac{
            {\range{n}{j}{s}{s}}
            {\pmva{s}}^{\!\trp}
        }{
            \lVert \pmva{s} \rVert^2
        }
    \!\bigg)\!\!
    \Bigg)
    \house{s}
        & \text{SB at $s$} 
        \\
        \!\Bigg(\!
        \jacobMVAsb
        \!-\!
        2
        \bigg(\!
            \frac{
                (\house{s'}{\range{n}{j}{s}{s'}})^{\!\trp}
                \pmva{s}
            }{
                \lVert \pmva{s} \rVert^2
            }
            \eye{}
            \!-\!
            \frac{
                \house{s'}{\range{n}{j}{s}{s'}}
                {\pmva{s}}^{\!\trp}
            }{
                \lVert \pmva{s} \rVert^2
            }
        \!\bigg)\!\!
        \Bigg)
        \house{s}
        & \text{DB $\MVApair{s}{s'}$} 
        \\
        \!\Bigg(\!
        \jacobMVAblock{s'}{s}
        \!-\!
        2
        \bigg(\!
            \frac{
                {\range{n}{j}{s'\!}{s}}^{\!\trp}
                \pmva{s}
            }{
                \lVert \pmva{s} \rVert^2
            }
            \eye{}
            \!-\!
            \frac{
                {\range{n}{j}{s'\!}{s}}
                {\pmva{s}}^{\!\trp}
            }{
                \lVert \pmva{s} \rVert^2
            }
        \!\bigg)\!\!
        \Bigg)
        \house{s}
        \house{s'}
        & \text{DB $\MVApair{s'\!}{s}$} \\
    \end{cases} \nonumber   
\end{aligned}
$
\end{adjustbox}
\end{equation*}

Partitioning the $(2\,S \times \Ncomponents)$ Jacobian matrices $\jacobM{\eta} = \big[{\jacobM{\eta,1}}\!^\trp \ \dots \ {\jacobM{\eta,S}}\!^\trp\big]^\trp$ mapping to all \glspl{surface} into $(2 \times \Ncomponents)$ matrices $\jacobM{\eta,s}$ mapping to a single \gls{surface}, with $\eta\! \in\! \{\dist{s}{s'}, \aoa{s}{s'}, \aod{s}{s'} \}$ denoting the respective local channel parameter,
we find the following Jacobian submatrix expressions:

Mapping to \gls{surface} position $\pmva{s}$, the Jacobians
\begin{align}
    \left[\jacobM{\delay,s}\right]_{\scriptscriptstyle :,\ncomponent} &= 
    \frac{\partial\ranget{n}{j}{s}{s'}\!^\trp}{\partial\pmva{s}}
    \frac{\partial\rangep{n}{j}{s}{s'}\!^\trp}{\partial\ranget{n}{j}{s}{s'}}
    \frac{\partial \delayx{j} }{\partial\rangep{n}{j}{s}{s'} }  
    \label{eq:jacobMdelay}
    \\
    &= 
    \jacobMVAblock{s}{s'} ~ \house{s'}\,\house{s} ~ \rotM ~ 
    \frac{\rangep{n}{j}{s}{s'}}{  \lVert \rangep{n}{j}{s}{s'} \rVert} \quad \in \realset{2}{1}     \,, \nonumber \\
        \left[\jacobM{\azimuthAoA,s}\right]_{\scriptscriptstyle :,\ncomponent}
    &= 
    \frac{\partial\range{n}{j}{s}{s'}\!^\trp}{\partial\pmva{s}}
    \frac{\partial\rangepTX{n}{j}{s}{s'}\!^\trp}{\partial\range{n}{j}{s}{s'}}
    \frac{\partial \azxAoA{j} }{\partial\rangepTX{n}{j}{s}{s'} }  \\
    &= 
    \jacobMVAblock{s}{s'}
    ~ \left(-\rotMn\right) ~ 
     \frac{1}{\rxTX^2 + \ryTX^2}
    \begin{bmatrix}
        -\ryTX \\
        \rxTX \\
    \end{bmatrix} \quad \in \realset{2}{1} \,,\nonumber \\
    \left[\jacobM{\azimuth,s}\right]_{\scriptscriptstyle :,\ncomponent} &= 
    \frac{\partial\ranget{n}{j}{s}{s'}\!^\trp}{\partial\pmva{s}}
    \frac{\partial\rangep{n}{j}{s}{s'}\!^\trp}{\partial\ranget{n}{j}{s}{s'}}
     \frac{\partial \azx{j} }{\partial\rangep{n}{j}{s}{s'} }  \\
    &= 
    \jacobMVAblockG{s}{s'} ~ \rotM ~ 
    \frac{1}{\rx^2 + \ry^2}
    \begin{bmatrix}
        -\ry \\
        \rx 
    \end{bmatrix} \quad \in \realset{2}{1} \, ,\nonumber 
\end{align}
map from distance, \gls{aoa}, and \gls{aod}, respectively. 
Note that $\jacobMVAblockG{s}{s'}\rotM\rangep{n}{j}{s}{s'}\!=\!\jacobMVAblock{s}{s'}\house{s'}\house{s}\rotM\rangep{n}{j}{s}{s'}$ in~\eqref{eq:jacobMdelay}.
Columns corresponding to the \gls{los} path or paths not involving surface $s$ are zero.

%